\documentclass[pra,aps,amssymb,amsmath,amsmath,showpacs,preprint]{revtex4-1}
 \usepackage{color}
\definecolor{Red}{rgb}{0.9,0.0,0.1}
\definecolor{Green}{rgb}{0.0,0.6,0.0}
\definecolor{Blue}{rgb}{0.0,0.0,0.6}

\usepackage{graphicx}
\usepackage{epstopdf}
\usepackage{natbib}
\usepackage{hyperref}
\usepackage{dcolumn}
\usepackage{bm}
\usepackage{dcolumn}
\newcolumntype{d}[1]{D{.}{.}{#1}}
\newcommand{\dk}{\partial_{\cal E}^2\!}
\newcommand{\bp}{\bm{p}}

\newcommand{\bps}{{\bm p}\,}
\newcommand{\bpns}{{\bm p}}

\newcommand{\qw}{\phantom{$-$}}

\newcommand{\bphi}{\bm{\varphi}}
\newcommand{\br}{\bm{r}}

\newcommand{\bP}{\bm{P}}

\newcommand{\ba}{\bm{a}}

\newcommand{\bU}{\bm{U}}
\newcommand{\bk}{\bm{k}}

\usepackage{hyperref}
\hypersetup{
pdftitle={QED calculation of the dipole polarizability of helium atom},
pdfsubject={Funder name: European Metrology Programme for Innovation and
Research, Funder ID: 10.13039/100014132, Grant number: QuantumPascal project 18SIB04
},
pdfauthor={M. Puchalski, K. Szalewicz, M. Lesiuk, and B. Jeziorski},
}

\newcommand{\Fkt}[1]{\,\mathsf {#1}}
\def\openone{\leavevmode\hbox{\small1\kern-3.3pt\normalsize1}}

\ifx\Tr\renewcommand{\Tr}{\Fkt{Tr}} 
\else\newcommand{\Tr}{\Fkt{Tr}}
\fi

\newcommand*{\cent}[1]{\multicolumn{1}{c}{$#1$}}

\usepackage{graphicx}
\usepackage{amsmath}
\usepackage{indentfirst}
  \def\v{\boldsymbol}
\def\k{\omega}

\def\pol{\alpha_{\rm d}}
\def\psiz{\psi_0}

\def\llangle{\big\langle}
\def\rrangle{\big\rangle}
\begin{document}
\title{QED calculation of the dipole polarizability of helium atom}

\author{\sc Mariusz Puchalski}
\affiliation{\sl  Faculty of Chemistry, Adam Mickiewicz University, Umultowska 89b, 61-614 Pozna\'n, Poland }
\author{\sc Krzysztof Szalewicz}
\affiliation{\sl  Department of Physics and Astronomy, University of Delaware, Newark, Delaware 19716, USA }
\author{\sc Micha\l \ Lesiuk and Bogumi\l\ Jeziorski}
\affiliation{\sl Faculty of Chemistry, University of Warsaw\\
Pasteura 1, 02-093 Warsaw, Poland}
\date{\today}

\begin{abstract}
The QED contribution to the dipole polarizability of the $^4$He atom was computed, including the effect of finite
nuclear mass. The computationally most challenging contribution of the second electric-field derivative of the
Bethe logarithm was obtained using two different methods: the integral representation method of Schwartz and
the sum-over-states approach of Goldman and Drake. The results of both calculations are consistent, although
the former method turned out to be much more accurate. The obtained value of the electric-field derivative of the
Bethe logarithm, equal to $0.048\,557\,2(14)$ in atomic units, confirms the small magnitude of this quantity found in
the only previous calculation [G. Łach, B. Jeziorski, and K. Szalewicz, Phys. Rev. Lett. 92, 233001 (2004)], but
differs from it by about 5\%. The origin of this difference is explained. The total QED correction of the order of α 3
in the fine-structure constant α amounts to 30.6671(1)$\cdot 10^{-6}$, including the 0.1822$\cdot 10^{-6}$ contribution 
from
the electric-field derivative of the Bethe logarithm and the 0.01112(1)$\cdot 10^{-6}$ correction for the finite nuclear
mass, with all values in atomic units. The resulting theoretical value of the molar polarizability of helium-4
is $0.517\,254\,08(5)\,$cm$^3$/mol with the error estimate dominated by the uncertainty of the QED corrections of
order $\alpha^4$ and higher. Our value is in agreement with but an order of magnitude more accurate than the result
$0.517\, 254\, 4(10)\,$cm$^3$/mol of the most recent experimental determination [C. Gaiser and B. Fellmuth, Phys. Rev.
Lett. 120, 123203 (2018)].
 \end{abstract}

\maketitle

\section{Introduction}
\label{sec:intro}

Accurate knowledge of the electric dipole polarizability  $\alpha_{\rm d}$ of helium is critical for the development of new  primary standards of temperature \cite{Moldover:16,Gaiser:17a,Rourke:19} (which is of importance due to the  new  definition of kelvin \cite{Fischer:19,Machin:19}),  and for novel realizations of pressure employing electrical \cite{Gugan:80,Gaiser:19a}, microwave \cite{Schmidt:07}, or optical methods \cite{Jousten:17,Hendricks:18}.  This significance of the polarizability  is a consequence of the direct linear relation 
 $(\varepsilon_{\rm r}-1)  k_{\rm B}T = 4\pi \alpha_{\rm d}\, p$  
connecting at low density the  relative electric permittivity $\varepsilon_{\rm r}$  (and consequently the refractive 
index $n$)  with the gas pressure $p$ and the thermodynamic temperature $T$.  The Boltzmann constant $k_{\rm B}$, 
appearing here,  is now fixed at $1.380649   \cdot\! 10^{-23}$ J/K.  
  Corrections to this linear relation,   depending  the second and  higher powers of density $\rho$, are small fo
helium \cite{Gavioso:16,Jousten:17} and can be determined with much lower relative accuracy than the targeted accuracy 
of $p$ or $T$.  Information about  an accurate  value of   $\alpha_{\rm d}$  is also essential  in experimental 
determinations of density and dielectric virial coefficients of rare gases using dielectric-constant gas thermometry  
\cite{Gaiser:19,Guenz:17}.  One may note that knowledge of the accurate value of the dipole polarizability of helium 
was employed  the experimental determinations of the value of the Boltzmann 
constant  \cite{Egan:17,Gaiser:17}, before this constant  was fixed by the new SI definition of kelvin~\cite{Fischer:18,Pitre:19}.   

For microwave \cite{Schmidt:07} and optical \cite{Jousten:17,Hendricks:18} methods,  the dependence of  $\alpha_{\rm d}$ on frequency $\omega$ is  relevant,  but for  helium the frequency dependent part of   $\alpha_{\rm d}(\omega)$ is small \cite{Piszczatowski:15} for experimentally  useful frequencies  \cite{Stone:04},
 and does not have to be known with high relative accuracy.  One  may also note that the index of refraction depends not 
only on $\alpha_{\rm d}(\omega)$ but also on the static magnetic susceptibility $\chi$ and, at the $10^{-7}$ level,  on 
other frequency dependent magnetic and  quadrupole contributions \cite{Pachucki:19}.  In this paper we  consider only  
the static dipole polarizability~$\alpha_{\rm d}$.

Since the helium atom is a very small system bound by electromagnetic forces, its properties, including the polarizability, can be computed with very high accuracy using the quantum electrodynamics (QED) theory.  The strong nuclear forces can be accounted for by the empirical values of 
 the nuclear mass and nuclear  charge radius. 
 The nuclear polarizability and effects of the  weak interactions give a completely negligible contribution 
to the atomic polarizability. 
The current status  of the QED  theory in the description of the helium atom 
has been recently examined in Ref.~\onlinecite{Pachucki:17}.  No relative discrepancies  
higher than $10^{-8}$ have been found \cite{Pachucki:17}  between the best theoretical calculations of transition energies and their most reliable experimental determinations.
In some cases  the agreement between theory and experiment reaches even  the $10^{-9}$ level \cite{Zheng:17}.  Since in thermal
metrology  the required relative accuracy is at most at the   
 $10^{-7}$ level,  one can be confident that the theory 
tested in Ref. \onlinecite{Pachucki:17}  is sufficient  for a metrology-useful prediction of the static  polarizability of helium.        

The nonrelativistic polarizability of helium $\alpha_{\rm d}^{(0)}$,  defined by the standard Schr\"odinger-Coulomb 
equation, can   be computed with accuracy limited only by the accuracy of the experimental value of the 
electron-to-nucleus mass ratio. The most accurate value of $\alpha_{\rm d}^{(0)}$ for $^4$He reported in the literature, 
1.383\,809\,986\,408(1) $a_0^3$, where  $a_0=\hbar^2/(m_{\rm e}e^2)$ is the atomic unit  of length,  has a relative 
error of  $10^{-12}$ (see Table I  in Ref. \onlinecite{Puchalski:16}).  The leading relativistic correction  to 
$\alpha_{\rm d}^{(0)}$,  being of the second order in the fine-structure constant $\alpha$  and  denoted by $\alpha_{\rm 
d}^{(2)}$,  can be computed using the Breit-Pauli Hamiltonian \cite{Bethe:75} and is also known with more than 
sufficient  accuracy.  Its value for  $^4$He is  $-$80.4534(1)$\cdot 10^{-6}$ $a_0^3$  \cite{Piszczatowski:15},  the 
reported uncertainty of $10^{-10}$ $a_0^3$  accounts for neglected terms of the order of $\alpha^2  (m_{\rm 
e}/m_{\alpha})^2$, where $m_{\alpha}$ is the nuclear mass.          

Calculation of the next correction, $\alpha_{\rm d}^{(3)}$,  of the order of $\alpha^3$, requires a field-theoretic,  QED treatment of the electron-electron  and electron-nucleus interaction that takes into account the effects of the electron self-energy and  the vacuum polarization. 
 The first calculation
of $\alpha_{\rm d}^{(3)}$ was reported  by Pachucki and Sapirstein  \cite{Pachucki:00a} in 2001. These authors assumed the infinite nuclear mass, i.e., considered  the nuclear-mass-independent  part   $\alpha_{\rm d}^{(3,0)}$ of  $\alpha_{\rm d}^{(3)}$,  and neglected the 
computationally demanding second electric-field derivative  $\partial^2_{\cal E} \ln k_0$    of the  so-called Bethe 
logarithm $\ln k_0$.  To estimate the uncertainty  of their calculation they assumed  that   $\partial^2_{\cal E} \ln 
k_0$ expressed in atomic units represents at most 10\% \  of the known  field-independent value of $\ln k_0$,  which  
translated into about 10\% \ error in   $\alpha_{\rm d}^{(3,0)}$. The complete calculation
of   $\alpha_{\rm d}^{(3,0)}$, including the effect of $\partial^2_{\cal E} \ln k_0$,  was reported in Ref. 
\onlinecite{Lach:04}. The obtained value of   $\partial^2_{\cal E} \ln k_0$,  equal to 0.0512(4) in atomic units,  
turned out to be about 
an order of magnitude smaller than  the estimate made by Pachucki and Sapirstein \cite{Pachucki:00a} and about two orders of magnitude smaller than the atomic value  of  $\ln k_0$.  The  $\partial^2_{\cal E} \ln k_0$ independent part of $\alpha_{\rm d}^{(3,0)}$   obtained by Lach et al. \cite{Lach:04}  agreed  well with the  calculations of  Pachucki and Sapirstein~\cite{Pachucki:00a}. 

The calculation of   $\partial^2_{\cal E} \ln k_0$ is computationally complex and error-prone 
since  it involves numerical treatment of divergent integrals, and since the final, unexpectedly small value of  $\partial^2_{\cal E} \ln k_0$ results  from cancellations of terms   much larger 
than $\partial^2_{\cal E} \ln k_0$. 
Therefore, it  is clear that   an independent confirmation of the results of Ref. \onlinecite{Lach:04} 
is needed. 
The main purpose of the present  paper is to perform a substantially more accurate calculation of  
$\partial^2_{\cal E} \ln k_0$  to verify the accuracy of the value obtained in Ref.     \onlinecite{Lach:04} and to 
obtain an improved    value of  $\alpha_{\rm d}^{(3,0)}$. 
To achieve this goal we   employed two different methods to compute  Bethe logarithms:   the modification of the integral representation  method of Schwartz \cite{Schwartz:61} proposed 
recently by Pachucki and Komasa \cite{Pachucki:04} and  the  sum-over-states method of 
 Goldman and Drake~\cite{Goldman:83,Goldman:84,Goldman:84a} modified by us to compute the second derivative of $\ln 
k_0$.  Another objective of this paper is to include the nuclear-mass-dependent  part $\alpha_{\rm d}^{(3,1)}$ of 
 $\alpha_{\rm d}^{(3)}$,   referred to 
as  the QED recoil correction.   By adding the computed values  
of  $\alpha_{\rm d}^{(3,0)}$  and  $\alpha_{\rm d}^{(3,1)}$, a definitive value of the $\alpha^3$ QED correction to the 
polarizability of helium will  become  available  
for metrological and other applications.   
 
The plan of this paper is as follows.  In Secs. II and III  we present calculations of  $\partial^2_{\cal E} \ln k_0$ 
performed using the integral representation and the sum-over-states methods, respectively.  Section  IV contains the 
description of the 
calculation of the   QED  recoil  correction to the polarizability of helium.  Finally, in Sec. V a summary of the 
obtained results is presented and the conclusion of this paper is formulated.  
The Appendix contains a  derivation of  a constant defining the 
 the asymptotic behavior of the integrand used in Sec. II to compute  $\partial^2_{\cal E} \ln k_0$.

Unless otherwise  stated,  atomic units are used throughout this paper. We   assume that 
$\alpha$=1/137.0359991,  $a_0=0.052917721$ nm  and that the mass of the $^4$He  nucleus equals
7294.2999536 $m_{\rm e}$. For the Avogadro number, we take  the new SI value of \
6.02214076$\cdot$10$^{23}$.

 \section{Integral representation   approach  to  the electric-field derivative of the Bethe logarithm }
\label{Sec_II} 

The formula for $\alpha_{\rm d}^{(3,0)}$ can be obtained by the electric-field differentiation of the  general expression for the $\alpha^3$ QED correction $E^{(3,0)}$  to the  
energy of two  electrons in a nondegenerate singlet state,   derived  
by  Araki  and Sucher   \cite{Araki:57,Sucher:58} in the 1950s.  In the compact, present-day  notation this formula can be written  
in the form (see e.g.,   Ref.  \onlinecite{Pachucki:17})
\begin{equation}
\begin{split} \label{E30}
 E^{(3,0)} = & \alpha^3\,\Big [\frac{8}{3}\,\Big( \frac{19}{30}-2\ln{\alpha}-\ln{k}_0 \Big)   {  D}_1  +       
   \Big({\frac{164}{15}+\frac{14}{3}\ln\alpha}\Big) { D}_2  - \frac{7}{6\pi}  \,  {  A}_2 \Big ],
 \end{split}
\end{equation}
where 
\vspace{-1ex}
\begin{eqnarray}\label{D1}
& {D}_1 &  =
\llangle{\psi }\vert{\delta^{3}(\v{r}_1)+\delta^{3}(\v{r}_2)}\vert{\psi }\rangle, \\ \label{D2}
& \     D _2 \ & =  \llangle{\psi }\vert\delta^{3}(\v{r}_{12})\vert{\psi }\rrangle ,  \\
 \label{P2}
&  A_2  &  =  \llangle{\psi }\vert P( r_{12}^{-3})\psi) \rrangle \equiv \lim_{a\to{0}}\llangle{\psi }\vert\theta({r_{12}-a})\,r_{12}^{-3} 
  +4\pi\,(\gamma+\ln{a})\,\delta^{3}(\v{r}_{12})\vert{\psi }\rrangle,  
\end{eqnarray}
with $\delta^3(\br)$ being  the three-dimensional Dirac distribution,  $\gamma$   the Euler-Mascheroni constant,   $\theta(x)$   the
Heaviside step function, and $\psi$    the ground-state eigenfunction  of the  nonrelativistic  electronic Hamiltonian  $H$ of the considered system.  
The quantity  $\ln{k_0}$, appearing also in Eq. (\ref{E30}),  is the Bethe logarithm defined as the  quotient 
\begin{equation}\label{BL1}
 \ln{k_0}=  
 \frac{\langle{\psi }\vert\,\v{p}\,{(H -E )}\ln{[2(H -E )]}\,\v{p}\,\vert{\psi }\rangle}
 {\langle{\psi }\vert\,\v{p}\,(H -E )\,\v{p}\,\vert{\psi }\rangle}\ ,
\end{equation}
where $E$  is the ground-state eigenvalue of $H$, i.e.,   $(H-E)\psi=0$, and  $\v{p}=\v{p}_1+\v{p}_2$ is the total momentum operator  for the electrons.   The numerator and the denominator   in Eq. (\ref{BL1}) will be denoted by 
${  N}$ and ${  D}$, respectively. One can show that  ${  D} =4\pi {D_1}$. 
In our case,  $H =H_0 + {\cal E}(z_1+z_2) $, where $H_0$  is the nonrelativistic electronic Hamiltonian  for the helium
atom and ${\cal E}(z_1+z_2) $ is  the perturbation due to a uniform static electric field ${\cal E}$ directed along the 
$z$ axis.  Thus, all quantities in Eqs. (\ref{E30})-(\ref{BL1}) depend on the electric-field strength 
${\cal E}$. In this and in the next section,  we assume that the nuclear mass is infinite and that $H_0$ contains only electronic kinetic energy.  

  Differentiating Eq. (\ref{E30})  twice with respect to ${\cal E}$ and reversing the sign, one obtains \cite{Pachucki:00a,Lach:04}         
\begin{equation}
\begin{split} \label{a30}
 \pol^{(3,0)} = & \ \  \alpha^3\,\Big[ -\frac{8}{3 }\Big({\frac{19}{30}-2\ln{\alpha}-\ln{k_0}}\Big)
           \partial^2_{\cal E} {   D}_1        
+ \frac{8}{3  } {  D}_1  \, \partial^2_{\cal E} \ln\! k_0    \\[1.5ex]
  & 
  - \,\Big({\frac{164}{15}+\frac{14}{3}\ln\alpha}\Big) \partial^2_{\cal E}D_2  +  \,\frac{7}{6\pi}  \,  \partial^2_{\cal E} A_2 \Big].
 \end{split}
\end{equation}
where all electric-field  derivatives  and the  quantities $\ln  k_0$  and $  D_1$    which are not differentiated
are   taken  at ${\cal E} =0$ .    

The evaluation of the derivatives $\partial^2_{\cal E}{D}_1 $,  $\partial^2_{\cal E} D_2$, and   $\partial^2_{\cal E} A_2$ is relatively easy and can be done using the   double-perturbation 
theory formula,  
\begin{equation} \label{E21}
 \partial^2_{\cal E} X=  4 \, \langle{\psiz}\vert  z  R_0  z  R_0\hat{X}  {\psiz}\rangle 
   + 2 \,\langle{\psiz}\vert z   R_0(\hat{X} - \langle \psiz | \hat{X} \psiz \rangle)R_0 z  {\psiz}\rangle, 
 \end{equation}
where $X$=$  D_1$, $D_2$, or $A_2$;  \ $\hat{X}$ stands   for  the operators  appearing in Eqs. (\ref{D1})-(\ref{P2});  
\  $z$=$z_1+z_2$, 
$\psiz$ is the  ground-state eigenfunction  of $H_0$,  i.e.,  $H_0 \psiz = E_0 \psiz$;  
 and  $R_0=(1-P_0)( H_0-E_0 +P_0)^{-1}$  is the  reduced   resolvent of $H_0$,  with
$P_0$ being the projection on $\psi_0$.

    To evaluate   $\partial^2_{\cal E} D_1 $,  $\partial^2_{\cal E} D_2$, and   $\partial^2_{\cal E} A_2$   via  Eq. (\ref{E21}),  we 
need two auxiliary functions:  the first-order function   $ R_0 z {\psiz}$ of natural $P$ symmetry
and the $S$-wave part  of the second-order function    $R_0zR_0z {\psiz}$. 
These auxiliary functions were represented using  the basis set of exponentially correlated Slater functions of the form 
\begin{equation}\label{Basis} 
{\tilde \psi}(\v{r}_1,\v{r}_2) = (1 + {\cal P}_{12})
\sum_{i=1}^{K } c_i\, Y(\v{r}_1,\v{r}_2)\, e^{-\xi_i r_1 - \eta_i r_2 - \nu_i r_{12} } \,,
\end{equation}
where  ${\cal P}_{12}$  exchanges   vectors $\v{r}_1$ and $\v{r}_2$ and  
$Y(\v{r}_1,\v{r}_2)$ is the angular factor equal to $z_1$ or 1 in the present  case. 
The linear and nonlinear parameters in Eq. (\ref {Basis}) were 
obtained by minimizing the  static form ($\omega =0$) of the Hylleraas  functional  
\begin{equation}\label{Hyll}
{\cal F}[\tilde{\psi}] = \langle \tilde{\psi} | H_0 - E_0 + \omega | \tilde{\psi} \rangle + 2 \langle \tilde{\psi} |h \rangle
\end{equation}
where the function $h$ is equal to $z \psi_0$ or  $z R_0  z \psi_0$.  
 The ground-state wave function $\psiz$ was  also represented  by  Eq. (\ref{Basis}). 
All nonlinear parameters $\xi_i$, $\eta_i$, and  $\nu_i$ were fully optimized  for bases with $K$ equal to 128, 256, 
and 512. 
 The results are shown in Table \ref{I}. 
  \begin{table}[ht]
  \caption{Mean values and their second electric-field derivatives
 obtained with the basis sets optimized in this paper. The values of $\sigma$   are  conservative error  estimates of 
the  values computed for $K=512$. They were  obtained by observing the  pattern of   convergence with increasing $K$ and 
by performing additional calculations 
with other basis sets.  \vspace{-3ex}}
  \label{I}
  \begin{center}
  \begin{tabular}{cd{4.11}d{4.11}d{1.11}d{3.11}d{2.10}d{3.8}}
     \hline \hline \\[-4ex]
  $K $ & \cent{D_1} &  \cent{\partial_{\cal E}^2 {D_1}} & \cent{D_2} & \cent{\partial_{\cal E}^2 D_2} & \cent{A_2} &  \cent{\partial_{\cal
 E}^2 A_2} \\[1ex]
  \hline
  128       & 3.620\,860\,71  & -5.168\,613\,9  & 0.106\,345\,341  & -0.394\,937\,6   & 0.989\,274\,6   & -2.573\,745\  \\
  256       & 3.620\,858\,67  & -5.168\,624\,4  & 0.106\,345\,364  & -0.394\,937\,4   & 0.989\,273\,9   & -2.573\,764  \\
  512       & 3.620\,858\,63  & -5.168\,624\,1  & 0.106\,345\,370  & -0.394\,937\,4   & 0.989\,273\,6   & -2.573\,766 \\
  $\sigma$  & 0.000\,000\,01  &  0.000\,000\,1  & 0.000\,000\, 001 &  0.000\,000\,1   &  0.000\,000\,2  &  0.000\,002  \\[1ex]
  \hline\hline
  \end{tabular}
  \end{center}
\end{table}

Inspecting the values collected  in Table \ref{I},  we see that our calculations of   $  D$,  $\partial^2_{\cal E}{   D 
}$,  $\partial^2_{\cal E} D_2$, and   $\partial^2_{\cal E} A_2$ are accurate    to  better than  1-ppm level.  Using the 
values obtained with the largest basis set and the best literature value \cite{Korobov:19} of the atomic Bethe 
logarithm 
$\ln k_0 = 4.370\,160\,223\,070\,3(3)$, 
 we find that the neglect of  $\partial^2_{\cal E} \ln k_0$, 
i.e., the   approximation used by Pachucki and Sapirstein  \cite{Pachucki:00a},
leads to the value of   $30.4738(1)\cdot 10^{-6}$  as an  approximation to  $\pol^{(3,0)}$. This value agrees very  well with the result of $30.474(1)\cdot 10^{-6}$ published  in  Ref.  \onlinecite{Pachucki:00a}.
 
The computation of  the electric-field derivative of $\ln k_0$ is substantially  more complicated  than the computation 
of expectation values   $D_1$, $D_2$, and $A_2$ and their electric-field derivatives. 
In this section we present the calculation  of  $\ln k_0$ using the integral representation method of Schwartz  \cite{Schwartz:61} in a computationally convenient  formulation  proposed by Pachucki and Komasa \cite{Pachucki:04}.   In this formulation,  the electric-field dependent Bethe  logarithm $\ln k_0$ is computed  as  the integral 
\begin{equation} \label{intlnk0}
 \ln  k_0 = \int_0^1   \frac{f(t)-f_0-f_2t^2}{{  D}\,t^3}\, dt ,
\end{equation} 
where   $f_0 = \langle\psi \vert \v{p}^2\psi \rangle $,  $f_2 =-2{D}$, and the function
$f(t)$ is defined by  
\begin{equation} \label{J}  f(t) = 
\omega J(\omega) = \omega \,  \langle\psi \vert  \v{p}\,  ({  H} - {  E} +\omega)^{-1} \v{p}\psi  \rangle 
\end{equation}  
with $\omega =(1-t^2)/(2t^2)$.    
%
 The denominator  $ D$  as well as the expectation values in  the definitions of  $J(\omega)$, $f_0$ and $f_2$  are assumed here to be obtained  
with the electric-field-dependent ground-state eigenfunction  $\psi$ of $H$.    
 Schwartz \cite{Schwartz:61} and Forrey and Hill \cite{Forrey:93}  developed the asymptotic, large-$\omega$ expansion  of $J(\omega)$ that   can  be transformed into the expansion of $f(t)$ at small $t$ which, up to the $t^4$ term,  takes the form \cite{Pachucki:04}
\begin{equation} \label{fexp}  
f(t)  \sim   f_{\rm exp}(t) = f_0 + f_2\, t^2 + f_3\, t^3 + f_{\rm 4 l}\,t^4  \ln t + f_{4}\,t^4 ,
\end{equation}
where $f_3 = 16{  D}$, \ $f_{\rm 4l} = 64 {  D}$ and $f_{4} = 2 {  D} \, (8 C_3 + 16 \ln 2  -1) $.
The constant $C_3$ determines the $\omega^{-3}$  term (equal to $4{  D}C_3\,\omega^{-3}$)  in the asymptotic expansion 
of $J(\omega)$.  The computation of $C_3$ and its electric-field derivative  $\partial^2_{\cal E}C_3$ is  discussed  in  
the Appendix.     
 
Performing the electric-field differentiation of Eq. (\ref{intlnk0}) and setting ${\cal E}=0$,  one obtains
\begin{eqnarray}\label{d2k1}
\partial_{\cal E}^2 \ln k_0 &=& \int_0^1 \! \frac{\partial_{\cal E}^2 f(t)-\partial_{\cal E}^2 f_0 -\partial_{\cal E}^2 f_2 \, t^2}{{  D}\,t^3} dt
 - \frac{\partial_{\cal E}^2 { D}}{{ D}} \ln k_0,     
\end{eqnarray} 
%
where  $\partial_{\cal E}^2 f_0 = \partial_{\cal E}^2 \langle \psi_0 \! \mid\!\! \v{p}^2 \psi_0 \rangle$ and   
$\partial_{\cal E}^2 f_2 =-2  \partial_{\cal E}^2{  D}$, whereas   ${   D}$ and  $\ln k_0$ on the right-hand side of 
Eq. (13) represent the atomic, field-independent values of  these quantities. Equation (\ref{fexp}) shows 
that the integrand  in Eq. (\ref{d2k1})  is finite at $t=0$ so  the integral is convergent. 
However, 
at small values of the argument $t$,  the function  $\partial_{\cal E}^2 f(t)$ is  very difficult to compute accurately using finite basis set expansions.
 Actually, when $\partial_{\cal E}^2 f(t)$, $\partial_{\cal E}^2 f_0$, and $\partial_{\cal E}^2 f_2$   are computed using a finite basis of the form of Eq. (\ref{Basis}), the  singularity of the integrand at $t=0$ is not canceled 
  and the integral   diverges.
 To circumvent this difficulty, the integral over $t$ 
was separated into two parts: part 1 from zero to $\epsilon \ll 1$ and part 2 from  $\epsilon$ to $1 $, with only  part 
2 computed using numerical values of $\dk f(t)$. P
art 1 was  obtained by approximating $\dk 
f(t)$ using  Eq. (\ref{fexp}) and its generalization involving higher powers of $t$. To reduce the contribution from part 1, it is convenient 
to subtract   $\partial_{\cal E}^2f_3\,t^3 +\partial_{\cal E}^2 f_{\rm 4l}\,t^4 \ln t +\partial_{\cal E}^2f_4\, t^4$ from the numerator in the integrand of Eq. (\ref{d2k1})  and integrate the counterterms analytically.  The resulting expression for  $\partial_{\cal E}^2 \ln k_0$  takes then the form
  \begin{eqnarray}\label{d2k2}
\partial_{\cal E}^2 \ln k_0 &=& \int_0^1 \! \frac{\partial_{\cal E}^2 f(t)-\partial_{\cal E}^2 f_{\rm exp}(t) }{{  D}\,t^3} dt +\frac{\partial_{\cal E}^2 f_4}{2D} - \frac{\partial_{\cal E}^2{   D}}{{   D}} \ln k_0, 
\end{eqnarray}   
where 
\begin{equation}\label{d2f4} 
\partial_{\cal E}^2f_{4} =    16\, {  D}\,\partial_{\cal E}^2C_3 +    2\,\partial_{\cal E}^2 {   D} \, (8 C_3 + 16 \ln 2  -1) .
\end{equation}
To  derive Eq. (\ref{d2k2}), use has been made of the fact that the integral over
$f_3 + f_{4{\rm l}}\, t \ln t$    accidentally  vanishes  for helium. The integrand $I(t)$ in Eq. (\ref{d2k2})  behaves at small $t$ as $ f_{5{\rm l}}  \, t^2 \ln t + f_5  \, t^2$ and for small $\epsilon$ gives a very small contribution to  $\partial_{\cal E}^2 \ln k_0$. Accurate computation of    $ f_{5{\rm l}}$ and  $f_5$   would be very difficult and was not attempted. 
Approximate values of these parameters  were  obtained by interpolating $I(t)$ for  
$0\! <\! t\! <\!  \epsilon$  using    a
    few  $t \geq \epsilon$   values of $I(t)$, see Eq. (17).  

From Eq. (\ref{d2k2}) we see that to obtain   $\partial_{\cal E}^2 \ln k_0$ we need 
(in addition to ${  D}$ and  $\partial_{\cal E}^2 D$)  accurate 
values of  $\partial_{\cal E}^2\langle\psi_0\vert \v{p}^2\psi_0 \rangle $,  $C_3$,   $\partial_{\cal E}^2 C_3$,  
and     $\partial_{\cal E}^2 f(t)$ for $t \geq \epsilon$ .  The computation of   
 $\partial_{\cal E}^2\langle\psi_0\vert \v{p}^2 \psi_0\rangle $  and   $\partial_{\cal E}^2 C_3$ was performed using Eq. (\ref{E21})  and the basis set of Eq. (\ref{Basis}).  The computation of   $\partial_{\cal E}^2 C_3$   and $C_3$  is somewhat intricate  since 
  matrix  elements  that have to be evaluated are more complex than the matrix elements of  $\bp$ or $\delta^3(\br)$ 
(see the Appendix for  details).  The results of these computations    
are displayed in Table \ref{II}. 

\begin{table}[ht] 
\caption{Parameters defining the  behavior of $f(t)$  at small $t $. See caption to Table I for the definition  of 
$\sigma$.}
\label{II}
\begin{center} 
\begin{tabular}{cd{2.16} d{5.16}d{4.10}d{4.8}}
    \hline \hline \\[-3ex]
$K \ \ \  \ \ $ &      \cent{  \langle\psi_0\vert \v{p}^2\psi_0 \rangle  } & \cent{ \partial_{\cal E}^2\langle\psi_0\vert \v{p}^2\psi_0 \rangle }  & \cent{C_3} &  \cent{\partial_{\cal E}^2 C_3} \\[1ex]
\hline 
  128 \ \ \ \       &       6.125\,587\,703\,817\,09 &  -9.012\,082\,333\,63    & 5.000\,826 & -0.049\,28  \\
256  \ \ \ \        &  6.125\,587\,704\,239\,64  &-9.012\,082\,339\,72  & 5.000\,634 & -0.052\,49 \\
512  \ \ \ \      &    6.125\,587\,704\,239\,93   &  -9.012\,082\,339\,74& 5.000\,624^a   & -0.052\,30 \\
$\sigma $ \ \ \ \ &     0.000\,000\,000\,000\,02&  0.000\,000\,000\,01& 0.000\,002 &0.000\,02 \\[0.6ex] 
\hline\hline
\end{tabular}
 \end{center}
$^a$ \footnotesize{In Eq. (29) of Ref. \onlinecite{Korobov:12},   Korobov uses  the value   5.000\,624\,87  without giving an uncertainty estimate.}
\end{table}
In view of the very strong cancellation between    $\partial_{\cal E}^2f(t)$  and  $\partial_{\cal E}^2\langle\psi_0\vert \v{p}^2\psi_0 \rangle$ at small $t$, it is important that the accuracy  of $\partial_{\cal E}^2\langle\psi_0\vert \v{p}^2\psi_0 \rangle$ is very high.  As shown in Table \ref{II} this quantity was 
computed with a relative  error  of $10^{-12}$.

The calculation $\partial_{\cal E}^2f(t)$ was done via the computation of   $\omega\, \partial_{\cal E}^2J(\omega)$ for  $\omega =(1-t^2)/(2t^2)$.  
 The  appropriate expression for $\partial_{\cal E}^2J(\omega)$ is obtained  by double  electric-field  differentiation  of Eq. (\ref{J}). The result of this differentiation  can be written in the form  
\cite{Lach:04}
 \begin{align}\nonumber
\ \ \  \partial^2_{\cal E}J(\omega) & = 4\, \langle\psi_0 | z\, R_0\,z\, R_0\,\bp  R(\omega)\,\bp  \psi_0\rangle  +  4\, \langle\psi_0 | z\, R_0\,\bp R(\omega)\,z  R(\omega)\,\bp  \psi_0\rangle \\ 
 & + 2\, \langle\psi_0 | z\, R_0\,\bp R(\omega)\,\bp  R_0\,z \psi_0\rangle +   2\, \langle\psi_0 | \bp R(\omega)\,z R(\omega)\,z  R(\omega)\,\bp  \psi_0\rangle \nonumber   \\  \label{d2J}
   &   -2\, \langle\psi_0 | z\, R^2_0\, z \psi_0\rangle  \, \langle\psi_0 | \bp R(\omega)\,\bp \psi_0\rangle  
  -2\, \langle\psi_0 | z\, R_0\, z \psi_0\rangle  \, \langle\psi_0 | \bp R^2(\omega)\,\bp \psi_0\rangle , 
 \end{align}
where  $ R(\omega)$=$(H_0-E_0+\omega)^{-1}$  is the frequency dependent resolvent of the  field-free Hamiltonian $H_0$. 
Some terms   in Eq. (\ref{d2J}) are singular at $\omega=0$,  
but these singularities as well as the $\omega$ independent parts cancel  so that $\partial^2_{\cal E}J(0)=0$ and, as a 
consequence, both  $\partial^2_{\cal E}f(t) $  and  the derivative of  $\partial^2_{\cal E}f(t) $ with respect to $t$ 
vanish at $t=1$.

  To evaluate  $\partial^2_{\cal E} f(t)$  via  Eq. (\ref{d2J}),  we can employ  the functions  $ R_0z\psiz$ and   $R_0zR_0z\psiz$  used to obtain $\partial^2_{\cal E}D$ but we also have to compute, for each value of $\omega$,   
several auxiliary functions: the first-order function  $R(\omega)\v{p} {\psiz}$ as well as  the scalar, pseudovector,  and tensor  components of the second-order functions  $ R(\omega)\v{p}R_0z {\psiz}$ and 
$R(\omega)z R(\omega)\v{p}{\psiz}$.  All these functions were computed variationally for each required value of $\omega$ 
using  appropriate  versions of the functional (\ref{Hyll}).  The trial functions $\tilde{\psi}$  were expanded using 
the basis set of Eq. (\ref{Basis}) with the angular factors  corresponding to the symmetry of the considered auxiliary 
function.  For the vector and pseudovector  functions we  set   
$Y(\v{r}_1,\v{r}_2) = x_1$ or  $z_1$ and $x_1z_2 -z_1x_2$, respectively.  For the functions of $D$ symmetry 
the basis consists of two parts  each containing $K$ terms: the first part with the angular  factor  $x_1z_1$  or $r_1^2 -3z_1^2$  and the second part with the  factors  $x_1 z_2+ z_1 x_2$  or $\v{r}_1\v{r}_2- 3z_1z_2$.
 For each value of $t$ on a grid of 100 points between 0.01  and 1.0 (and a few  additional  points below 0.01), full optimizations of all nonlinear parameters were performed for three successively increasing basis sets labeled by  the integers $K$\,=\,128, 256,  and 512  which  specify also the size of the basis used  to expand  
$\psiz$.   

In Table 
\ref{III} we show the  basis set convergence of the integrand  $I(t)$  in Eq.~(\ref{d2k2}) for small  values of $t$.
 It is seen that the convergence, very good at $t > 0.005$, deteriorates dramatically for  small  values of $t$.   
At $t=0.002$,  the value of  $I(t)$ is not accurate enough to be used in numerical integration. This is shown in Table \ref{IV} where we list the values of the integral of $I(t)$ from $\epsilon$ to 1 computed with our two largest basis sets.   
The integral from 0.005 to 1   turns out to be sufficiently accurate and we have chosen $\epsilon = 0.005$ to separate  the integration range  in Eq.~(\ref{d2k2}) into the ``small   $t$" and ``large $t$" parts.  Using  $\epsilon$ larger than 0.005 gives more accurate values 
of the large $t$ integral (cf. Table \ref{IV}),   but is not advantageous since, as shown in  Table \ref{V},  the error 
of the whole calculation is determined 
by the interpolation error in the range  $t\!<\!\epsilon$  (performing the integration using  every second point we verified that  the error of our numerical integration procedure is smaller than 10$^{-8}$ and therefore negligible compared to other error sources).

 \begin{table}[ht]
\caption{Basis set convergence of the integrand $I(t)$ in Eq.~(\ref{d2k2}) for small  values of  $t$.  
 $K$ denotes the  basis set size used to represent $\psiz$ and the auxiliary functions. Extrapolated results were obtained assuming exponential decay of error. The uncertainty $\sigma$ is defined as the difference of the two preceding rows.}\vspace{0ex}
\label{III}
\begin{center}
\begin{tabular}{cd{5.10}d{4.10}d{2.10}d{3.10}d{3.10} }
    \hline \hline \\[-3ex]
$K$    & \cent{I(0.002)} &  \cent{I(0.005)} & \cent{I(0.01)} & \cent{ I(0.02)} & \cent{I(0.03)}\\[1ex]
\hline
128          &  -21.91782900  &  0.14613471 &  2.22617850   & 9.34234812  & 20.52412502  \\
256          &  -2.78361795   & 0.49908734 &  2.30166902  &9.34931045 &20.52630907   \\
512          & 0.03568909     &  0.50676476 & 2.30195231   &9.34932498  & 20.52631193   \\
extrp.       &0.52288119      & 0.50693547&2.30195337  & 9.34932501  &20.52631194    \\
$\sigma$   & 4.9 \cdot10^{-1} & 1.7\cdot10^{-4} & 1.1\cdot10^{-6}   & 3.0\cdot10^{-8} & 3.8\cdot10^{-9}   \\
\hline
\end{tabular}
\end{center}
\end{table}  

\begin{table}[ht]
\caption{Integral of $I(t)$ from $\epsilon$ to 1 computed with our two largest basis sets.   The uncertainty $\sigma $ is defined as the difference of the two preceding rows.}\vspace{0ex}
\label{IV}
\begin{center}
\begin{tabular}{cd{5.10}d{4.10}d{2.10}d{3.10}d{3.10} }
    \hline \hline \\[-3ex]
$K$    & \cent{\epsilon=0.002} &  \cent{\epsilon=0.005} & \cent{\epsilon=0.01} & \cent{\epsilon=0.015} & \cent{\epsilon=0.02}\\[1ex]
\hline
256          &  65.32851048  &65.32854759&  65.32840496   &65.32799881 &65.32720487  \\
512          & 65.32856319   & 65.32854787 & 65.32840479    &65.32799878  & 65.32720488    \\
$\sigma$   & 5.3\cdot10^{-5} & 2.8 \cdot10^{-7} & 1.7\cdot10^{-7}  & 3.4\cdot10^{-8} & 2.3\cdot10^{-9}   \\
\hline
\end{tabular}
\end{center}
\end{table}  

The integral from zero to $\epsilon$ was obtained analytically by interpolating $I(t)$ with the function   
\begin{equation} \label{fit1}
 \tilde{I}(t) = \sum_{k=2}^n (a_{k }\,t^k\ln t + b_k\,t^k) 
\end{equation} 
using our best (extrapolated) values of $I(t)$ for $t=\epsilon$ and for $2n-3 $ next higher values  of $t$.  
The results of this integration    are shown  in Table \ref{V} as   a function of $n$  together with the corresponding
values  of $ \partial^2_{\cal E}\ln k_0$ obtained from Eq.~(\ref{d2k2}) using our  best values  of $   D $ and 
 $ \partial^2_{\cal E}D$  (from Table \ref{I}),  of  $C_3$ and  $ \partial^2_{\cal E}C_3$  (from Table \ref{II}),  
and of the large $t$ integral  (from Table~\ref{IV}).  
One should note that the obtained values of   $\partial^2_{\cal E}\ln k_0$ are more than three orders of magnitude smaller 
than the individual terms in Eq.~(\ref{d2k2}).

\begin{table}[h]
 \caption{Dependence   of the integral  of  $I(t)$ from zero to $\epsilon =0.005 $ and of  the value of 
$\partial_\mathcal{E}^2\ln k_0$ on the length $n$ of the fit function of Eq. (\ref{fit1}). 
 For the $t\ge \epsilon$  integral  we took  $65.32854787$ (cf. Table~\ref{IV}).  } 
\begin{center}
\label{V}
 \begin{tabular}{ccc}\hline\hline
 $n$ &\ \ \   $\int_0^{\epsilon} I(t) dt$  &\ \ \   $\partial_\mathcal{E}^2 \ln k_0$ \\
 \hline
 2 & \ \ \  0.00001736    &  \ \ \   0.04855859   \\
 3 &\ \ \   0.00001636   &\ \ \   0.04855759  \\
 4 &\ \ \   0.00001608   &\ \ \   0.04855731 \\
 5 &\ \ \   0.00001599  & \ \ \  0.04855722   \\
\hline\hline
 \end{tabular}
\end{center}
\end{table} 
Table \ref{V} shows that the integral from zero to $\epsilon$ 
is very small but its relative accuracy is not high.  From the observed convergence pattern we  can infer 
that  the value of this  integral amounts to 0.0000160(14) with the  uncertainty conservatively estimated  by the
total spread of   values shown in Table \ref{V}.    Taking into account the error  estimations for both integration regions,
we  find  that the   value of $ \partial^2_{\cal E}\ln k_0$ obtained using the integral representation method is  0.0485572(14).  This value differs by about 5\% \ from the value 0.0512(4) reported in Ref.~\onlinecite{Lach:04}.  The origin of this difference is discussed in Sec. III.

  \vspace{-2ex}


\section{Sum-over-states approach to   the electric-field derivative of the Bethe logarithm}\vspace{-2ex}

To resolve the discrepancy between the values of $\dk \ln k_0$ obtained in Sec.~\ref{II} and 
in Ref.~\onlinecite{Lach:04}, we performed  computations using the sum-over-states approach
\cite{Goldman:83,Goldman:84,Goldman:84a}.  In  this approach,   the numerator 
 ${N}$
in Eq. (5) is represented by the spectral expansion in terms of the eigenfunctions   $\psi_n$ 
of the excited states of the Hamiltonian $H$,
\begin{equation}\label{N1}
{N} = \sum _n \,  \omega_n \ln(2\omega_n)  \, |\langle \psi_0 |\bp \psi_n \rangle|^2,
\end{equation}
where   $\omega_n$ are the excitation energies.  
In practice, an expansion in terms of pseudostates   
diagonalizing  $H$ in an appropriately chosen basis set  is used \cite{Goldman:83}.  Although the pseudostate  expansion  is converging  extremely slowly  (it is on the verge of divergence  \cite{Goldman:00}), it has been successfully applied   \cite{Bhatia:98,Drake:99,Yan:03,Yan:08}, also in the acceleration gauge  \cite{Korobov:04, Zhong:13},
 to accurately compute electric-field-free values of $\ln k _0$ .  In this section we present the application of  this method to compute   $\dk \ln\!  k_0$ for the ground state of the  helium atom in   a  static electric field  ${\cal E}$.  
 
To cope with the extremely slow convergence of the pseudostate  expansion, we use a parameter $L>0$ which  attenuates the importance of highly excited states and enables us to   control   the convergence rate.   
 Using the  integral representation of  $\ln\omega_n$   
\begin{equation}\label{tlogt}
  \ln  \omega_n =  \ln(1+L) -   \ln\!\left(1 +\frac{L}{\omega_n}\right) +   (\omega_n-1) \int_L^{\infty}
 \frac{d \omega}{(\omega +\omega_n)(\omega +1)},
\end{equation}
one can show that 
${N}$ can be written in the form  
\begin{equation}\label{N2}
{N} = {N}_L  +   D  \ln(2L+2) 
                           + \int_L^{\infty}g(\omega) d \omega ,
\end{equation}
where
\begin{equation}\label{NL}
{N}_L = -\sum _n \,  \omega_n \ln\!\left(1 +\frac{L}{\omega_n}\right)  \, |\langle \psi_0 |\bp \psi_n \rangle|^2
\end{equation}
and
\begin{equation}\label{g}
g(\omega)  =   \omega J(\omega)  - \langle \psi_0 |\bpns^2\, \psi_0 \rangle   +  \frac{  D}{\omega+1}.
\end{equation}
  One may note that the  modification of the original approach of Goldman-Drake, 
 as defined  by Eqs. (\ref{N2})-(\ref{g}),  bears close  resemblance  to the approach used  by Korobov 
\cite{Korobov:12,Korobov:19} (see also Ref. \onlinecite{Korobov:12a}).  

When  the  energies $\omega_n$ of the excited states are large  (much larger than $L$)  the successive contributions in the summation   in 
Eq. (\ref{NL}) decrease with $n$ as $L|\langle \psi_0 |\bp \psi_n \rangle|^2 $.  This  should be compared with  the  
 $\omega_n  \ln  \omega_n   \, |\langle \psi_0 |\bp \psi_n \rangle|^2$  decrease of  terms in Eq. (\ref{N1}).  One can thus expect that 
the convergence of the summation  in the expression for ${N}_L$ will be  faster than the convergence of the series in  Eq. 
(\ref{N1}).   When $L$ is sufficiently large,  the last term in Eq.~(\ref{N2}) is small and can be easily computed using the large-$\omega$ asymptotic expansion of  $g(\omega)$  
\begin{equation}\label{gexp}
g(\omega)  =    g_3 \, \omega^{-3/2} 
  +g_{4\rm l}\, \omega^{-2} \ln\omega   +g_4 \,\omega^{-2}    + g_5\,  {\omega^{-5/2}}  + \cdots  ,
\end{equation}
where the    coefficients 
$g_3 = 4\sqrt{2}\,D $, \  $ g_{4\rm l}\,=  -8\,D, $ \,  and  $g_4= (4C_3 -1) \,D $ can be obtained by changing the variable in the expansion  of Eq.  (\ref{fexp}),
or directly from the work of Schwartz~\cite{Schwartz:61}.  Forrey and Hill \cite{Forrey:93} derived an expression for $g_5$ but this expression
is too complex to evaluate in practice. 

Carrying out the $\omega$  integration  in  Eq. (\ref{N2}) using the first three   terms  in the  asymptotic expansion of $g(\omega)$    
and adding the result to the first two terms in this  equation, one  obtains the following expression for $ \ln  k_0$:
\begin{equation}\label{BL0}
 \ln  k_0 =    \ln  k_0(L) + {\cal R}_L,
\end{equation}
where 
 \begin{equation}\label{BL}
 \ln  k_0(L) =  \frac{{N}_L } {D}  +  \ln (2L+2) +   8\sqrt{2} L^{-1/2}   - 8 L^{-1}\, {\ln L}   +   (4\,C_3 -9) {L^{-1}} 
\end{equation}
and  ${\cal R}_L$ is the error resulting from truncating the asymptotic series  of Eq. (\ref{gexp}).  
We know from the work of Forrey and Hill \cite{Forrey:93}    that    ${\cal R}_L$ 
vanishes  with increasing  $L$ as
\begin{equation}\label{err}    
{\cal R}_L =    C_4\,  L^{-3/2}  + C_5\, L^{-2} \,{\ln L}   +  C_6\, {L^{-2}}  + {\cal O}(  L^{-5/2}) .
\end{equation}
Knowing this error formula, one can perform the  extrapolation  of   $ \ln  k_0(L)$ and obtain an improved value  
of  $ \ln  k_0 $  by solving a small system of linear equations.
 
In view of Eq. (\ref{BL}), the second electric-field derivative of $\ln k_0(L)$  is given by  the expression
\begin{equation}\label{d2log}
\partial^2_{\cal E}\ln\!k_0 (L) = \frac{1}D\left[\,\partial^2_{\cal E}{N}_L -  \frac{{N}_L}D\, \partial^2_{\cal E}D\,\right] 
 + \frac{4}{L} \,\partial^2_{\cal E}C_3.
\end{equation}
%
%
The derivative of the error $\partial^2_{\cal E}{\cal R}_L$  has   the same  large-$L$ behavior as ${\cal R}_L$  so  that
 $\partial^2_{\cal E} \ln \!k_0 (L)$ can be extrapolated in the same way as $ \ln \!k_0(L)$ using Eq.  (\ref{err}).
%
%

Since the intermediate wave  functions $\psi_n$  of the pseudostates and the excitation energies  $\omega_n$  in Eq. (\ref{NL})  depend on the electric field ${\cal E}$, the  differentiation 
of $ {N}_L $  with respect to ${\cal E}$  is much more difficult than the differentiation of $D$ or $  \langle\psi\vert \v{p}^2\psi \rangle  $.  A suitable sum-over-states 
expression for   $\partial^2_{\cal E}{N}_L$ can be obtained from the formula 
\begin{equation}\label{dNL3}
\partial^2_{\cal E}{N}_L = \int_0^L \omega\,\partial^2_{\cal E} J(\omega) d\omega   - L\,\partial^2_{\cal E}\langle \psi_0 |\bp{\!}^2 \psi_0 \rangle 
\end{equation}
resulting  from  Eqs. (\ref{N2}) and (\ref{g}).  Using Eq. (\ref{d2J})
 and  noting that terms diverging linearly with $L$  are eliminated  
with the help of Eq. (\ref{E21}), one finds that    $\partial^2_{\cal E}{N}_L $ can  be written as the  sum of six contributions
 \begin{equation}\label{sum6}
\partial_{\cal E}^2 {{N}_L} = I_A + I_B+I_C+I_D+I_E+I_F,
\end{equation}
defined by  
\begin{eqnarray}\label{IA} 
I_A &=& -4\sum_n \lambda(\omega_n) \langle\psi_0 | zR_0zR_0\bp  \psi_n \rangle  \langle \psi_n|\bp   \psi_0\rangle ,\\
\label{IB}
I_B &=& -2\sum_n \lambda(\omega_n)  |\langle\psi_0 | z\,R_0\,\bp \psi_n\rangle|^2  ,\\
\label{IC}
I_C &=& 2 \langle\psi_0 | z\,R^2_0\, z \psi_0\rangle  \sum_n \lambda(\omega_n)|\langle\psi_0 | \bp\psi_n\rangle|^2, \\\label{ID}
I_D &=& -2\, \langle\psi_0 | z\,R_0\, z \psi_0\rangle  \sum_n \kappa(\omega_n) |\langle\psi_0 | \bp \psi_n\rangle |^2, \\\label{IE}
I_E &=&  4\, \sum_{k}\sum_{n}  \gamma(\omega_k,\omega_n)\,\langle\psi_0 | z\,R_0\,\bp\psi_k\rangle\langle \psi_k|z \psi_n\rangle\langle\psi_n|\bp  \psi_0\rangle,\\\label{IF}
I_F &=& 2 \sum_l\sum_k\sum_n  \phi(\omega_l,\omega_k,\omega_n) \,\langle\psi_0 | \bp \psi_l\rangle\langle\psi_l|z\psi_k\rangle\langle\psi_k|z \psi_n\rangle\langle\psi_n|\bp  \psi_0\rangle, 
\end{eqnarray}
where
 \begin{eqnarray}\label{gt}
\lambda(t)& =& t \ln\!\left(1+\frac{L}{t}\right), \\[1ex]
\label{h}
\kappa(t)& = & \ln\!\left(1+\frac{L}{t}\right) - \frac{L}{L+t},\\[1ex]
 \label{Gst}
\gamma(s,t)&=&\frac{\lambda(s)-\lambda(t)}{s-t},\\[1ex]
\label{Frst}
\phi(r,s,t)& =&- \frac{\lambda(r)}{(r-s)(r-t)}   - \frac{\lambda(s)}{(s-t)(s-r)} -\frac{\lambda(t)}{(t-r)(t-s )}.
\end{eqnarray}
Equations (\ref{Gst}) and (\ref{Frst})  are valid when all arguments  $r$, $s$, and $t$ are different. 
If $t = s$ then $\gamma(s,s) = \kappa(s)$. This case is very unlikely, however, since the states $\psi_k$  and $\psi_n$ 
 in Eq. (\ref{IE}) are of different parity. 
  The function  $\phi(r,s,t)$ is symmetric in its arguments. This may be used to simplify 
somewhat the summations in Eq. (\ref{IF}).  When only two arguments are equal, for instance $r$ and $t$  ($\psi_l $ and $\psi_n$ are of the same parity),  one  obtains  
\begin{equation}\label{Frrt} 
\phi(r,s,r) =      \frac{r}{(r-s)^2} \,\left[  \ln \left(1+\frac{L}{r}\right) -   \ln \left(1
+\frac{L}{s}\right)\right]  +\frac{L}{(L+r)(r-s)}.
\end{equation} 
In an unlikely case when all arguments are equal ($\psi_k$ must be of different parity than that of $\psi_l$ and  $\psi_n$),   
one finds
\begin{equation}\label{Frrr} 
\phi(r,r,r) =      \frac{L^2}{2r(L+r)^2}.
\end{equation} 
%

To obtain the final formula for the analytic second   derivative of   the Drake and Goldman expression 
 for $\ln k_0$, we have to  eliminate the logarithmic divergencies  in the square brackets of   
Eq.   (\ref{d2log})  by taking the limit   $L\rightarrow \infty$. This   is not entirely straightforward  since the  logarithmic divergencies in  the individual components of $\partial^2_{\cal E}{N}_L$,  given by Eqs.  (\ref{IA})-(\ref{IF}),   must be   isolated and shown to  cancel against appropriate counterterms 
resulting from $({N}_L/ {D} )\, \partial^2_{\cal E} {D}  $.  
 
To identify these counterterms, we replace  ${N}_L $ in Eq. (\ref{d2log}) by  the large-$L$ estimate   
\begin{equation}\label{NLoverD}
 {N}_L  =   {N}   -  {D} \ln(2 L) +
{\cal O}( L^{-1/2}),
\end{equation}
resulting from Eqs. (\ref{N2}) and (\ref{gexp}),   and 
write the difference  in the square brackets in  Eq.~(\ref{d2log}) as 
\begin{equation}\label{diff}
 \partial^2_{\cal E}{N}_L -  \frac{{N}_L} {D} \,\partial^2_{\cal E} {D}   = 
 \partial^2_{\cal E}{N}_L     +   \partial^2_{\cal E} {D}  \ln L -   ( \ln k_0 -\ln 2)\, \partial^2_{\cal E} {D}  +
{\cal O}( L^{-1/2}).
\end{equation}
   The derivative $\partial^2_{\cal E} {D} $ is calculated in practice  using  
 the relation $D=4\pi D_1$ with   $D_1$ given by the right-hand side  of Eq.  (\ref{D1}), but to obtain the   
counterterms needed to cancel the logarithmic divergence of  the individual contributions to  $\partial^2_{\cal E}{N}_L 
$ [cf. Eqs. (\ref{IA})-(\ref{IF})] we  differentiate  the  expression 
$\langle \psi_0 \mid \bp (H-E_0)\,\bp \psi_0 \rangle$ also defining $D$. The second derivative of this expression  at ${\cal E}$=0  is
 \begin{eqnarray} \nonumber
  \partial^2_{\cal E} {D}  & =& 2 \langle \partial^2_{\cal E}\psi_0 \mid \bp (H-E_0)\,\bp \psi_0 \rangle 
 - \langle \psi_0 \mid\! \bpns^2 \psi_0 \rangle   \partial^2_{\cal E}E_0\\ \label{d2D}
&+ &2  \langle \partial_{\cal E}\psi_0 \mid \bp (H-E_0)\,\bp \partial_{\cal E}\psi_0 \rangle + 4 \langle \partial_{\cal E}\psi_0 \mid \bp z\,\bp \psi_0 \rangle ,
  \end{eqnarray}
where  $\partial_{\cal E}\psi_0$=$\,- R_0z \psi_0$,   
  $\partial^2_{\cal E}\psi_0$\,=\,$2R_0zR_0 z \psi_0$$-$$\langle \psi_0 \mid\! zR^2_0 z  \psi_0 \rangle \, \psi_0$,
and $\partial^2_{\cal E}E_0$=$\,- 2\langle \psi_0 \mid\! zR_0 z  \psi_0 \rangle $, are the appropriate derivatives of the wave function  and the energy. Inserting these derivatives 
into Eq. (\ref{d2D}),  one finds that $ \partial^2_{\cal E} {D} $ can be written as a sum of the following five terms
\begin{equation} \label{D5}
\partial^2_{\cal E} {D}   =     {D}_A + {D}_B+ {D}_C+ {D}_D+ {D}_E ,
\end{equation}  \vspace{-1ex} 
 where  \vspace{0ex} 
\begin{eqnarray} \label{DA}
 {D}_A & =&   4\sum_n  \omega_n  \langle\psi_0 | zR_0zR_0\bp\! \psi_n \rangle  \langle \psi_n|\bp   \psi_0\rangle ,\\
 \label{DB}
 {D} _B  & =&   2\sum_n  \omega_n  |\langle\psi_0 | zR_0\bp\! \psi_n \rangle|^2 ,\\
 \label{DC}
 {D} _C  & =&  - 2 \langle\psi_0 | z\,R^2_0\, z \psi_0\rangle  \sum_n  \omega_n |\langle\psi_0 | \bp\psi_n\rangle|^2,\\
 \label{DD}
 {D} _D & =&   2 \langle\psi_0 | z\,R_0\, z \psi_0\rangle  \,   \langle\psi_0 |  \bpns^2\psi_0\rangle ,\\
 \label{DE}
 {D} _E  & =&   -4  \langle\psi_0 | zR_0\,\bp z \, \bp    \psi_0\rangle    .
\end{eqnarray}
Let us now consider the logarithmically divergent terms in Eqs.  (\ref{IA})-(\ref{IE}). 
 To isolate them we need the following large-$L$  estimates
\begin{eqnarray}\label{gL}
\lambda(t)& = &t \ln L  -t\ln  t  + {\cal O}\left(L^{-1}\right)\\
 \label{hL}
\kappa(t)& = & \ln L - \ln t -1 + {\cal O}\left(L^{-1}\right) .
\end{eqnarray}
Inserting Eqs. (\ref{gL}) and   (\ref{hL}) into Eqs. (\ref{IA})-(\ref{IE}),
  it is easy to see that all  terms proportional to $\ln L$  cancel exactly against 
the second term on the right-hand side of Eq.~(\ref{diff}). More specifically, the $\ln L$ component of $I_X$ 
  cancels against $ {D} _X \ln L $, where $X$=$A,B,C,D,E$ [cf. Eqs.~(\ref{DA})-(\ref{DD})].  
What remains after these cancellations   is the sum of contributions given by    Eqs.~(\ref{IA})-(\ref{IE}) 
  in which the factors $\lambda(t)$ and  $\kappa(t)$  are replaced 
by $  -t \ln t $ and  $  -\ln t -1 $, respectively, and the $\gamma(s,t)$ factor is replaced by 
\begin{equation}\label{Ginfst}
\gamma_{\infty}(s,t)= -\frac{s \ln s -t \ln t}{s-t}
\end{equation} 
for $s\neq t$ and  by $\gamma_{\infty}(t,t) = - \ln t -1$, when $s$=$t$. 

To finish the discussion of the $L \rightarrow \infty$ limit, we  still have to consider the contribution from the $I_F$ term 
of Eq. (\ref{IF}), which is finite at large $L$.  One can easily show that the $L \rightarrow \infty$ limit of
the   factor $\phi(r,s,t)$, denoted by $\phi_{\infty}(r,s,t)$, is given by
\begin{equation}\label{Finfrst} 
\phi_{\infty}(r,s,t) = \frac{r \ln r}{(r-s)(r-t)}   + \frac{s \ln s}{(s-t)(s-r)} +\frac{t\ln t}{(t-r)(t-s )},
\end{equation} 
when all arguments  $r$, $s$, and $t$ are different, and by
\begin{equation}\label{Finfrrt} 
\phi_{\infty}(r,r,t) =      \frac{t\,(\ln t -\ln r)}{(r-t)^2} \,   +\frac{1}{ r-t },
\end{equation} 
 \begin{equation}\label{Finfrrr} 
\phi_{\infty}(r,r,r) =      \frac{1}{2r}.
\end{equation} 
when two of them or all three are equal. 

Summarizing, the final formula for the second derivative of the Goldman-Drake expression for the Bethe logarithm is  
\begin{equation}\label{final}
\partial^2_{\cal E}{\ln k_0} = \frac{1}{D}  (G_A +G_B +G_C +G_D+G_E+G_F) 
- (\ln k_0 - \ln 2)\, \frac{\partial^2_{\cal E}D}{D} , 
\end{equation}\vspace{-6ex} 

where \vspace{-5ex} 

\begin{eqnarray} \label{NA}
G_A & =&   4\sum_n  \omega_n  \ln \omega_n \,  \langle\psi_0 | zR_0zR_0\bp  \psi_n \rangle  \langle \psi_n|\bp   \psi_0\rangle ,\\
 \label{NB}
G_B & =&    2\sum_n  \omega_n \ln \omega_n \,  |\langle\psi_0 | zR_0\bp  \psi_n \rangle|^2\\
 \label{NC}
G_C& =&  - 2 \langle\psi_ | z\,R^2_0\, z \psi_0\rangle  \sum _n \,  \omega_n \ln(\omega_n)  \, |\langle \psi_0 |\bp \psi_n \rangle|^2, \\
 \label{ND}
G_D& =&  2\, \langle\psi_0 | z\,R_0\, z \psi_0\rangle \, \sum_n (1 +\ln \omega_n) |\langle\psi_0 | \bp \psi_n\rangle |^2,\\
 \label{NE}
G_E& =&   4\, \sum_{k}\sum_{n}  \gamma_{\infty}(\omega_k,\omega_n)\,\langle\psi_0 | z\,R_0\,\bp\psi_k\rangle\langle \psi_k|z \psi_n\rangle\langle\psi_n|\bp  \psi_0\rangle, \\
\label{NF}
G_F &=& 2 \sum_l\sum_k\sum_n  \phi_{\infty}(\omega_l,\omega_k,\omega_n) \,\langle\psi_0 | \bp \psi_l\rangle\langle\psi_l|z\psi_k\rangle\langle\psi_k|z \psi_n\rangle\langle\psi_n|\bp  \psi_0\rangle. 
\end{eqnarray} 
All components in the expression for $\partial^2_{\cal E}{\ln k_0}$ are finite,  but  substantial cancellations of individual terms
can occur and 
  the final value of  $\partial^2_{\cal E}{\ln k_0}$ is expected to be at least two orders of magnitude smaller than the individual contributions in Eq. (\ref{final}). 

Since ${N}_L =0$ when $L=0$,   one can think of deriving Eq.  (\ref{final})  by differentiation of  Eq.~(\ref{N2}) setting  $L=0$.  The individual integrals resulting then from the application of Eq. (\ref{d2J}) are divergent at infinity and require 
regularization 
that in practice is the same as the one used by us to derive Eqs.   (\ref{final})-(\ref{NF}). 
Thus,  $L$ can be viewed as a regularization parameter needed to derive the second derivative of the Goldman-Drake
expression for the Bethe logarithm. 
  

    The problem of finding a  basis to represent  pseudostates needed  to accurately  compute  
the quantities defined by  Eqs.  (\ref{NA})-(\ref{NF})  presents a serious   challenge.   Unless $L$ is very small, the same problem appears  in calculating  the quantities defined by    Eqs.  (\ref{IA})-(\ref{IF}).   
The difficulty stems from the fact that pseudostates  with extremely high energies are required to converge  the logarithmic sums of the form 
of Eq. (\ref{N1})  or Eq. (\ref{NL}) and, also,  from the  additional flexibility needed to describe the polarization by the external electric field. 
 To  obtain a suitable basis,  we followed  the procedure employed by Korobov~\cite{Korobov:00,Korobov:04}  in highly 
accurate calculations of Bethe logarithms 
for the ground and excited states of the helium atom. In his method,  the parameters $\xi_i$, $\eta_i$ and $\nu_i$ 
defining the basis functions  of Eq. (\ref{Basis}) are distributed stochastically   within 
one or several three-dimensional boxes while the positions and sizes of these boxes are determined by minimizing the Hylleraas functional of Eq. (\ref{Hyll}) setting $\omega=0$ and replacing the inhomogeneity function $h$  by $ h_1=     (z_1 r_1^{-3}  +z_2 r_2^{-3})\psi_0$. The singular behavior of $h_1$  at $r_i \rightarrow 0$ increases the flexibility of the basis at small $r_i$ which is needed to represent pseudostates with very high energies.  The inhomogeneity function $h_1$ was used by us to optimize bases of  natural $P$ symmetry. 
To optimize bases of  $S$ and $D$ symmetry,  we used the same Hylleraas functional but with the 
inhomogeneity $h$ replaced by  the  $S$ and $D$ part, respectively,  of the function 
$ h_2=  (z_1 r_1^{-3}  +z_2 r_2^{-3})R_0z\psi_0$. To optimize bases  of   $P^{\rm e}$ symmetry, the inhomogeneity  $h_3 =  ( x_1 r_1^{-3} R_0z_2  -  z_2 r_2^{-3} R_0x_1)\psi_0$ was used.

The basis set for pseudostates of natural $P$ symmetry needed to evaluate ${N}_L$ via Eq. (\ref{NL}) 
was  constructed as follows. We start with the {\em primary box }   $[A_1,A_2] \times [B_1,B_2]\times [C_1,C_2]$ 
with a uniform stochastic distribution of $K_0$ triples of real exponents $ \xi_i$, $\eta_i$, and $\nu_i $. This box 
defines  $K_0$   basis functions. Then, following  the ideas presented in  Refs. \onlinecite{Korobov:00,Korobov:04}, we 
build a set of {\em secondary   boxes } $ [\tau^kA^{ \prime }_1,\tau^kA^{  \prime }_2] \times [B^{ \prime }_1,B^{ \prime 
}_2]\times [C^{ \prime  }_1,C^{ \prime }_2]$, $k=0$$,\ldots,$$14$, where
 $\tau$=$A_2/A_1$ and where
$A^{ \prime }_1, A^{  \prime }_2, B^{ \prime }_1, B^{ \prime }_2, C^{ \prime  }_1, C^{ \prime }_2 $ are  parameters subject to  nonlinear optimization together with the  primary box parameters  $A_1,  A_2,  B_1, B_2, C_1,  C_2 $.  In each secondary box,  we distribute stochastically $n_k K_0/25$  basis functions, where $n_k$=10,8,6,5,4,3,3,2,2,2,1,1,1,1,1  for $k=0,\ldots,14$, respectively.  
The parameters $ \xi_i$, $\eta_i$, and $\nu_i $   were always  constrained by the conditions 
$\xi_i + \eta_i > \sqrt{2 I}$,
$\eta_i + \nu_i >\sqrt{2 I}$, and $\nu_i +\xi_i > \sqrt{2 I}$,
  where $I$ is the ionization
potential of helium.  This
ensures that the basis  functions  fall off
sufficiently rapidly when $r_1, r_2  \to \infty$ to   represent  a bound state.
If a randomly generated basis function fails to fulfill these conditions, it
is rejected and another one is generated.

  To represent pseudostates of $P$ symmetry, we used 3$K_0$  basis  functions defined by 12 nonlinear parameters.   The   bases with $K_0$ = 100, 200, and 400 were optimized. To represent $\psi_0$, we used a single box  with  $K_0$ basis functions
and box parameters determined by minimizing  the 
ground-state energy.
  The  helium atom energies  obtained using bases with $K_0$ = 100, 200, and 400 terms were only   9$\times$10$^{-10}$,  11$\times$10$^{-12}$, and 5$\times$10$^{-14}$ above 
the    accurate ground-state energy of the helium atom  \cite{Schwartz:06,Nakashima:07,Aznabaev:18}. 

Using the  pseudostates  obtained with the basis sets optimized as described above,    we evaluated ${N}_L$  via Eq. (\ref{NL}) for $L = 100, 200, 500, 1000$.  The resulting values of $\ln k_0(L)$, cf. Eq. (\ref{BL}),    are shown in  Table~\ref{VI}. It is seen that the convergence with increasing  $L  $  is very slow, as expected from  the error estimate of Eq. (\ref{err}),  and that  the basis set convergence is also slow, 
deteriorating appreciably with the increase  of $ L$.  The extrapolation to $L  = \infty$ based on the error estimate of Eq. (\ref{err}) is, however,  quite effective 
  reducing  the error of $\ln k_0$ by three orders of magnitude compared to the $L=1000$ value given in Table~\ref{VI}. 
Specifically, applying  Eq. (\ref{err}) for $L = 100, 200, 500, 1000$, neglecting the ${\cal O}(L^{-5/2})$ terms, and solving for  the 
unknown   variables $\ln k_0$, $C_4$, $C_5$, and $C_6$ we obtain $\ln k_0 = 4.370\,162\,1 $ when  $K_0$~=~400. This 
value has the  relative error of $ 5 \cdot 10^{-7}$ comparing to the  best available value   
 \cite{Korobov:19} and is significantly more accurate than the results
of the first two applications of the Schwartz method \cite{Schwartz:61,Baker:93}.     
\begin{table}[ht]
\caption{$L$-dependence of the approximate Bethe logarithm $\ln k_0(L)$ for helium. }
\begin{center}
\begin{tabular}{c c  ccc}
    \hline \hline
$K_0/ L$   &   \ \  100 &\ \ \  200 &\ \ \  500 &\ \ \  1000\\
 \hline
100   &  \ \  4.410\,654\,707 &\ \ \  4.385\,346\,797 &\ \ \  4.374\,273\,575 &\ \ \  4.371\,701\,275 \\
200   &  \ \ 4.410\,629\,878 &\ \ \  4.385\,315\,192 &\ \ \  4.374\,232\,108 &\ \ \  4.371\,652\,027 \\
400  &   \ \ 4.410\,629\,718 &\ \ \  4.385\,314\,986 &\ \ \  4.374\,231\,829 &\ \ \  4.371\,651\,683 \\
 \hline
\hline
\end{tabular}
\label{VI}
\end{center}
\end{table}

\begin{table}[ht]
\caption{  $L$ dependence of $\partial^2_{\cal E}\log k_0(L)$, see Eq.   (\ref{d2log}). 
For  $\partial^2_{\cal E}C_3$ we assumed the value $-$0.05230, see Table  \ref{II} and  Appendix.}
\begin{center}
\begin{tabular}{ccccc}
    \hline \hline
\ \ \ $K_0/ L$\ \ \  & 50 & \ \   100 & \ \ \   200 & \ \     500 \\
\hline
100 & 0.041\,523 &\ \ \  0.044\,121 &\ \ \  0.049\,995 & \ \ \   0.050\,930   \\
200 & 0.044\,327 &\ \ \  0.046\,739 &\ \ \ 0.048\,492 & \ \ \   0.049\,303  \\
400 & 0.044\,476 &\ \ \  0.046\,882 & \ \ \ 0.048\,075 & \ \ \   0.048\,617   \\
\hline\hline
\end{tabular}
\label{VII}
\end{center}
\end{table}

To calculate the second electric-field derivative of the Bethe logarithm,    we   need also  
bases   of scalar  $S$,  pseudovector  $P^{\rm e}$,  and  natural $D$ symmetry. 
The specific composition of these bases was as follows.
For the $S$ symmetry, we used seven  boxes. The first box,  containing $K_0/2$ functions,  was   the same as  optimized earlier in the calculations of the ground-state wave function  $\psi_0$.  The second box was optimized using the modified Hylleraas functional   and also contains $K_0/2$ functions. The remaining five boxes had exponentially growing sides $[\tau^nA^{\prime}_1,\tau^nA^{\prime}_2]$, $k=0,\ldots, 4$ with $\tau = A_2/A_1$  defined by the parameters $A_1$ and $A_2$ optimized for  the second box.  These boxes contain $n_k K_0/25$ basis functions,
where $n_k=7,6,5,4,3$ for $k=0,\ldots, 4$. In this way,  by optimizing 12 nonlinear parameters, we have generated the total of $2K_0$  scalar functions. 
 
For  pseudostates of  $P^{\rm e}$   and $D$ symmetry,  we used six boxes. The primary boxes  contained  $K_0$, and $3/2 K_0$ basis functions in the case  of the $P^{\rm e}$ and $D$ symmetry, respectively.  The remaining five boxes had
exponentially increasing sides as for the $S$ symmetry.  These five boxes contained  $K_0$, and $5/2 K_0$ basis functions for the $P^{\rm e}$   and $D$ symmetry, respectively, distributed  proportionally in the same    way as in the case 
of the last five, exponentially growing   boxes of $S$ symmetry.  In total, we stochastically generated  $2 K_0$ basis functions of   $P^{\rm e}$  symmetry and $4 K_0$ functions of  $D$ symmetry. In each case 12 nonlinear parameters were optimized. 
Bases for the  first-order functions $R_0z \psi_0$ and  $R_0p_z \psi_0$ and for the second-order function
 $R_0zR_0z\psi_0$  ($S$ symmetry only) contained $K_0$ elements  and were obtained from a single box, optimized
using appropriate Hylleraas functionals.  
 
Using the  bases  optimized for  $K_0$ = 100, 200, and 400, we evaluated  the $L$ dependence of 
$\partial^2_{\cal E}\ln\!k_0 (L)$ [see Eq. (\ref{d2log})]  for $L = 50, 100, 200, 500$. The results are shown in  
  Table~\ref{VII}.  It is seen that the convergence both in $K_0$ and in $L$ is much slower than in the case of 
$\ln k_0(L)$.  This is due to the loss of at least two digits in the subtraction in Eq. (\ref{dNL3}) and to  the much   increased basis set sensitivity of the components of Eq. (\ref{sum6}) 
compared to the already hard to converge summation in Eq. (\ref{NL}). The slowest convergence occurs in computing 
 the $I_B$ contribution of Eq. (\ref{IB}), which determines the final accuracy of    $\partial^2_{\cal E}\ln\!k_0 (L)$.  In fact,  the results for $L$=1000  were not accurate  enough to perform 
a reliable extrapolation and are not shown in  Table~\ref{VII}.  Also the values of the limit $L=\infty$ obtained from Eq. (\ref{final}) were very inaccurate   
and are not reported.

 Employing  the values of  $\dk \ln k_0(L)$  obtained with  $L = 50, 100, 200, 500$,  and the error formula of Eq. 
(\ref{err}), we find that the extrapolated values of $\dk \ln k_0$  are  0.04924 and 0.04875 when bases with  $K_0$=200 
and 400, respectively, are used. From these values one can infer that  the accurate value of  $\dk \ln k_0$ is smaller 
than 0.00487, in disagreement with the result of  Ref.  \onlinecite{Lach:04}. Based on the convergence pattern observed 
by us,  it  is very difficult to assign a reliable uncertainty to the value of $\dk \ln k_0$  resulting from  our 
sum-over-states calculation. We estimate  that this uncertainty is no worse than about 0.005 (i.e. about 1\%) and that  
our sum-over-states value  of $\dk \ln k_0$ amounts to  0.0487(5). This value  differs by 5\%\ from the value published 
in Ref.  \onlinecite{Lach:04}, but  is in perfect agreement with the value obtained by us in Sec. II using the Schwartz 
method. It is clear that in the case of  polarizability calculation   the Schwartz method is much more accurate (since 
the nonlinear optimizations are performed 
for each value of the frequency $\omega$) but the Goldman-Drake  approach can be used as an independent check of the 
result obtained using  the Schwartz method. 
 
We made some effort to explain the difference  (of about 5\%)  between the results of our calculations  (obtained using two different methods)  and the result 
of Ref.~\onlinecite{Lach:04} obtained by an application of the original version of the Schwartz method.  We found that  
the observed disagreement  has  three sources:  ($i$) the  omission of the singular,  $\psi_0$ contribution to the 
resolvent $R(\omega)$ in Eq. (\ref{d2J})  for    $\partial^2_{\cal E}  J(\omega)$ 
   used  in Ref.~\onlinecite{Lach:04} [the singularity $\omega^{-1}$ and the $\omega$-independent terms cancel out in 
the final expression  for  $\partial^2_{\cal E}  J(\omega)$,
so this contribution  is small],  ($ii$) the insufficiently accurate value of $\partial^2_{\cal E}\!\left<\psi_0\vert 
\bp^2\psi_0 \right>$  used in Ref. \onlinecite{Lach:04}  to evaluate the integral defining  
$\partial^2_{\cal E}\!  \ln k_0$,  and ($iii$) the fact that the value of $\partial^2_{\cal E} C_3 $ employed in Ref. 
\onlinecite{Lach:04}  was incorrect since it was computed from an incomplete formula, missing the explicit 
electric-field contribution $\partial^2_{\cal E} C_3^{(2)} $ derived in the Appendix of the present paper.

\section{QED recoil correction}

The theory of the nuclear mass dependence of the $\alpha^3$ QED correction  for two-electron systems has been given  by 
Pachucki in Ref. \onlinecite{Pachucki:98}. The expressions derived 
in this reference have been applied for the first time in Ref. \onlinecite{Pachucki:00} for the lowest $S$ states of the 
helium atom and subsequently for other excited states of helium \cite{Pachucki:17} and  helium like 
ions~\cite{Yerokhin:10}, as well as for  the low-lying states  of   lithium \cite{Puchalski:08}, beryllium 
\cite{Puchalski:14},  and  boron \cite{Maass:19} atoms. The leading correction $E^{(3,1)}$, of the order  of $1/M \equiv 
m_{\rm e}/m_{\alpha}$,  can be written  as the sum of  three contributions $E^{(3,1)}_{\rm R1}$,
$E^{(3,1)}_{\rm R2}$, and $E^{(3,1)}_{\rm R3}$. The first two represent the change  linear in $1/M$   of the  ingredients in Eq. (\ref{E30})  that  results from  adding to $H$  the nuclear kinetic energy operator 
 $  \bP^2/(2m_{\alpha})$  corresponding to the recoil momentum  $ \bP=-(\bp_1 +\bp_2)$.  
 The first contribution, $E^{(3,1)}_{\rm R1}$,   accounts for the effect of  
 $  p_1^2/(2m_{\alpha})+ p_2^2/(2m_{\alpha})$.  
It can be  obtained by  scaling
  Eq.~(\ref{E30}) with  the  reduced mass  $ \mu/m_{\rm e}  \approx 1- 1/M$,   resulting in  
\begin{equation}\label{EI}
 E^{(3,1)}_{\rm R1}  =  \   \frac{1}{M}  \left( -3E^{(3,0)} +
2{\cal E}\frac{\partial }{\partial\cal E}E^{(3,0)} +  \alpha^3 \,  \frac{8}{3} D_1  - \alpha^3\, \frac{14}{3} D_2\right), 
\end{equation}
where the second term in the  parentheses is a consequence of the electric-field  dependence of the scaled wave function 
$\mu^6\psi(\mu \br_1,\mu \br_2,\mu^{-2} { \cal E})$,   while the last  two terms originate  from the $\ln\mu$ dependence of the Bethe logarithm $\ln k_0$  \cite{Pachucki:00} and from the  $\mu^3\!\ln\mu^{\!-1} $  scaling of the Araki-Sucher term $A_2$~\cite{Yerokhin:10}.  
 The second  contribution,  $E^{(3,1)}_{\rm R2}$, is due  to the mass polarization term  $H_{\rm MP}=\bp_1\bp_2/m_{\alpha}$ and requires new calculations.   It  has the form   
 \begin{equation} 
\begin{split} \label{EII}
  E^{(3,1)}_{\rm R2}  = & \   \alpha^3  \frac{1}{M}\,\Big[ \frac{8}{3 }\Big({\frac{19}{30}-2\ln{\alpha}-\ln{k_0}}\Big)
           \partial_M \!{   D}_1        
- \frac{8}{3  } {  D}_1  \, \partial_M\! \ln\! k_0    \\[1.5ex]
  & 
 + \,\Big({\frac{164}{15}+\frac{14}{3}\ln\alpha}\Big) \partial_M\!D_2  -  \,\frac{7}{6\pi}  \,  \partial_M\! A_2 \Big],
 \end{split}
\end{equation} 
where $\partial_M$ denotes the derivative with respect to $1/m_{\alpha}$ when only the mass polarization term  $H_{\rm MP}$ is added to $H $.     The third contribution  is a  generalization of the Salpeter correction known for the hydrogen atom \cite{Salpeter:52}.  It has the form \cite{Pachucki:98}
\begin{equation} \label{EIII}
   E^{(3,1)}_{\rm R3}  =  \    \alpha^3 \frac{4}{M}\, \Big[ \Big(
-\frac{2}{3 } \ln \alpha  +\frac{62}{9} -  \frac{8}{3}\ln\!k_0\Big)\, D_1   - \frac{7}{6\pi} A_1
 \Big] ,
              \end{equation} 
  where  
\begin{equation}\label{A1}
A_1    =  \llangle{\psi }\vert P( r_{1}^{-3}) +  P( r_{ 2}^{-3})\vert  {\psi }\rrangle 
 \end{equation}
with the distribution  $P(r^{-3})$ defined by Eq. (\ref{P2}). 
  
When ${\cal E}=0$, the evaluation of  $ E^{(3,1)}_{\rm R1}$ and $ E^{(3,1)}_{\rm R3}$  is  no more difficult  than the evaluation 
of $E^{(3,0)}$. To evaluate $ E^{(3,1)}_{\rm R2}$,  we  need also the derivatives  $  \partial_M \!{D}_1$,    
$\partial_M \!{D}_2$, 
 $\partial_M \!{A}_2$, and  $\partial_M\! \ln\! k_0 $.  The first three of them can be easily obtained from the double perturbation  
theory expression $  \partial_M \!{   X}  =-2  \langle \psi \vert \hat{X} R_0 H_{\rm MP} \psi\rangle$,
 where  $ \hat{X}$  stands for the operators  appearing in Eqs. (\ref{D1})-(\ref{P2}).  Since $1/M$ is very small,  
these derivatives can also be obtained with sufficient accuracy  using the finite difference method. Analytic evaluation 
of the derivative  $\partial_M\! \ln\! k_0 $ is nontrivial. It has been performed for the first time by Pachucki and 
Sapirstein~\cite{Pachucki:00}. Currently the most accurate value of $  \partial_M\! \ln\! k_0 = 0.0943894(1)$ has been 
reported by Yerokhin and Pachucki  \cite{Yerokhin:10}. A somewhat less accurate  value  of $   \partial_M\! \ln\! k_0 = 
0.09438(1)$ has been obtained by Drake and Goldman \cite{Drake:99} using the finite difference
method.  Using the result from the former reference and the finite difference  calculation of  the remaining derivatives,   we found that 
$E^{(3,1)}$ = -5.12993$\cdot 10^{-9}$, in agreement  with the value  $-5.129925 \cdot 10^{-9}$  reported in Ref.~\onlinecite{Yerokhin:10}.   

 Performing   electric-field differentiation  of Eqs. (\ref{EI})-(\ref{EIII}), setting ${\cal E}\!=\!0$, and reversing 
the sign,  we find  
\begin{equation}\label{a31}
 \alpha_{\rm d}^{(3,1)} =  \alpha_{\rm R1}^{(3,1)} + \alpha_{\rm R2}^{(3,1)} + \alpha_{\rm R3}^{(3,1)}, 
\end{equation}
 where
\begin{equation}
\begin{split}  \label{R1}  
 \alpha^{(3,1)}_{\rm R1}  = & \     \frac{1}{M}\,\left(\alpha_{\rm d}^{(3,0)}         -\alpha^3\,
  \frac{8}{3}\, \partial^2_{\cal E}D_1 +\alpha^3\, \frac{14}{3}\, \partial^2_{\cal E} D_2\right), \phantom{ MMMMMMMMMM}
\end{split}
\end{equation}  
 \vspace{-6ex}

\begin{equation}
\begin{split}  
   \label{R2}
  \alpha^{(3,1)}_{\rm R2}  = & \   \alpha^3  \frac{1}{M}\,\Big[-  \frac{8}{3 }\Big({\frac{19}{30}-2\ln{\alpha}-\ln{k_0}}\Big)
           \partial_M  \partial^2_{\cal E}\!{   D}_1     -\,\Big({\frac{164}{15}+\frac{14}{3}\ln\alpha}\Big) \partial_M \partial^2_{\cal E}\!D_2   
   \\[1.5ex]
   &   +  \,\frac{7}{6\pi}  \,  \partial_M \partial^2_{\cal E}\!A_2  
+ \frac{8}{3  }\partial^2_{\cal E}  {  D}_1  \, \partial_M\! \ln\! k_0 
+ \frac{8}{3  }\partial_M\! {  D}_1  \,\partial^2_{\cal E}  \ln\! k_0
+ \frac{8}{3  } {  D}_1  \, \partial_M\!\partial^2_{\cal E}  \ln\! k_0\Big],
 \end{split}
\end{equation} \vspace{-3ex}

\begin{equation} \label{R3}
   \alpha^{(3,1)}_{\rm R3}  =  \    \alpha^3 \frac{4}{M}\,   \Big[\Big(  \frac{2}{3 } \ln \alpha  -\frac{62}{9} +
     \frac{8}{3}\ln\!k_0 \,\Big) \partial^2_{\cal E} D_1 +     \frac{8}{3} \, D_1\,\partial^2_{\cal E}\! \ln\!k_0 +\frac{7}{6\pi}\partial^2_{\cal E} A_1   \Big].
                \end{equation} 

  Equation (\ref{R1}) can also be obtained by performing the reduced mass scaling of Eq. (\ref{EI})  and observing that 
$\partial^2_{\cal E}\! \ln\!k_0$ scales  as $\mu^{-4}$ with the reduced mass $\mu$.
The first four  terms in   the square brackets of  Eq. (\ref{R2}) can be obtained by performing the  $ \partial_M$ differentiation of the approximate expression for 
$\alpha_{\rm d}^{(3,0)}$ used by Pachucki and Sapirstein \cite{Pachucki:00a}. 
Since, as found in Ref. \onlinecite{Lach:04} and confirmed in the present paper,  the  derivative  $\partial^2_{\cal 
E}\! \ln\!k_0 $ neglected by Pachucki and Sapirstein  is very small, we employed  the same approximation and neglected 
the last two terms  in the square brackets in Eq. (\ref{R2}). 
Actually, we know the contribution of the penultimate term, containing the product  
  $\partial_M\! {  D}_1  \,\partial^2_{\cal E}  \ln\! k_0$. This contribution equals to $ -1.3\cdot 10^{-12}$ and is completely negligible. 
The contribution of the last term can be estimated assuming that  the $\partial_M$ derivative of $\partial^2_{\cal E}  
\ln k_0$ is of the same order of magnitude as $\partial^2_{\cal E} \ln k_0$  (the  $\partial_M$  derivatives appear to 
be always smaller than or of the same order of magnitude as the differentiated quantities (see Table 1 of Ref. 
\onlinecite{Pachucki:00});    the same holds for the $\partial_M$ derivatives of $\partial^2_{\cal E} D_1$, 
$\partial^2_{\cal E} D_2$,  and $\partial^2_{\cal E} A_2$).   Making this assumption, we find that 
the neglected contribution of $\partial_M\partial^2_{\cal E}  \ln k_0$ is of the order of   $10^{-11}$ and is negligible compared to other contributions to the recoil correction. 
 This justifies the   Pachucki-Sapirstein approximation in evaluating   $ \alpha_{\rm d}^{(3,1)}$. 
To compute the $\partial_M\!$ derivatives of the expectation values,  we  used the finite difference method 
and our largest basis set,  $N=512$,
developed to obtain  the derivatives shown in Table  \ref{I}. We   have  
 found  that   $\alpha^{(3,1)}_{\rm R1}=   0.00484 $, $\alpha^{(3,1)}_{\rm R2}=0.00087 $, $\alpha^{(3,1)}_{\rm R3}=0.00541 $, and  that the whole QED recoil correction  $\alpha^{(3,1)}_{\rm d}$  is equal to 0.01112(1), with all values in the units of  $10^{-6}a^3_0$. The assumed uncertainty results from a conservative estimate of the neglected  electric-field derivatives of $\ln\!k_0$


 
  \section{Summary  of  the  Results and Conclusions}

We performed calculations of the main, $\alpha^3\ $QED contribution to the static polarizability of helium including 
the hard-to-compute  electric-field dependence of the Bethe logarithm and the finite nuclear mass (recoil) effects.   
This work 
complements earlier studies of the leading relativistic correction \cite{Cencek:01,Pachucki:00}, relativistic recoil effects  \cite{Piszczatowski:15}, and the QED correction in  the  infinite nuclear mass approximation \cite{Pachucki:00,Lach:04}.    
Our calculations of the second electric-field derivative of the Bethe logarithm  $\partial^2_{\cal E}  \ln\! k_0$, 
performed using  the  integral representation method of Schwartz ~\cite{Schwartz:61} (see Sec.~\ref{II}), confirm the 
very small value of this quantity found in Ref.  
\onlinecite{Lach:04}.  
 However,  the  value of 
 $\partial^2_{\cal E}  \ln\! k_0$ obtained by us,
 equal to 0.0485572(14), is smaller than  the value of Ref.~\onlinecite{Lach:04}, equal to 0.0512(4), by about six 
times the error estimate given in Ref.  \onlinecite{Lach:04}.  To resolve this discrepancy,   we performed (see Sec. 
\ref{III}) calculations of  
 $\partial^2_{\cal E}  \ln\! k_0$ using a different method based on the direct summation of 
the spectral representation of  $\partial^2_{\cal E}  \ln\! k_0$ in terms of pseudostates, along the lines  suggested by 
Goldman and Drake~\cite{Goldman:83}  and Korobov~\cite{Korobov:12}.  The result of this  second calculation,  equal to  
0.0487(5),  
   is  consistent with the result of the calculation using the integral representation method  of Schwartz but is  
inconsistent with the result  of Ref. \onlinecite{Lach:04}.

\begin{table}[h] 
\caption{Static polarizability of helium-4  (in $a_0^3$ unless otherwise noted)  including relativistic and QED 
corrections.  
The reported uncertainties are estimated based on the convergence in basis sets, except as marked.
When no error bar is given,  the last digit is certain. \vspace{2ex}}
\label{BD2}

\begin{tabular}{c l l l} \hline\hline  
 & contribution & \multicolumn{1}{c}{value} \\
 \hline
\qw  &  nonrelativistic          & \qw$1.383\,809\,986\,4^a$ & \\
&     $\alpha^2$ relativistic               &   $-0.000\,080\,359\,9^a $ \\
&    $\alpha^2/M   $   relativistic recoil \qw  &   $-0.000\,000\,093\,5(1)^b$ \\
& $\alpha^3$   QED    $ -\,\,\partial^2_{\cal E}\ln k_0$  term        & \qw$0.000\,030\,473\,8 $ \\
      &   $\partial^2_{\cal E}\ln k_0$  term       & \qw$0.000\,000\,182\,2 $ \\
   &   $\alpha^3/M$   QED recoil       & \qw$0.000\,000\,011\,12(1)^c $ \\
        &$\alpha^4$  QED                 & \qw$0.000\,000\,56(14)$$^{d}$ \\
      &   finite nuclear size      & \qw$0.000\,000\,021\,7(1)^{e} $ \\
      &   total                    & \qw$1.383\,760\,78(14)$ 
  
  \\ \hline 
&molar polarizability 
$\frac{4\pi}{3}\alpha_{\rm d}N_A$         & \qw$0.517\, 254\,08(5)^{f,g}$ \\
  &   experiment, Ref.~\onlinecite{Gaiser:18}
                                  & \qw$0.517\, 254\,4(10)^f$ \\
   
\hline \hline
\end{tabular}
\begin{flushleft}
$^a$ Ref. \onlinecite{Puchalski:16}.\\
$^b$ The uncertainty accounts for the included of terms  of the order of $1/M^2$ and  of higher    order~\cite{Piszczatowski:15}.\\
$c$ The uncertainty due to the neglect of the mixed derivative 
$ \partial_M\!\partial^2_{\cal E}  \ln\! k_0$ in Eq. (\ref{R2}).\\
$^d$  The uncertainty accounts for  an incomplete calculation  of the $\alpha^4$ QED correction to \qw polarizability, see Ref. \onlinecite{Piszczatowski:15}. \\
$^e$  Ref. \onlinecite{Piszczatowski:15}.\\
$^f$ In cm$^3$/mole. \\
$^g$ Using the nonrelativistic polarizability of the $^3$He atom, equal to 1.38401218(1) \cite{Puchalski:16} and scaling  \qw  the  recoil corrections with the mass ratio of    1.32711 one finds  that the molar polarizability of \qw  helium-3 is 0.517 329 65(5)  cm$^3$/mole. 
\end{flushleft}
\end{table}
After including the contribution of $\partial^2_{\cal E}\!  \ln k_0$, the total value of the $\alpha^3$ QED correction to the polarizability of helium in the infinite nuclear mass approximation amounts to 
\mbox{30.6560(1)$\cdot 10^{-6}$ $a_0^3$}. We   derived a  formula for the    
  correction to this value due to the finite nuclear mass (the QED recoil correction). In evaluating this formula we 
neglected $\partial^2_{\cal E}  \ln k_0$ and the  mass-polarization effect on $\partial^2_{\cal E}  \ln k_0$, given by 
the mixed derivative $\partial_M\partial^2_{\cal E}  \ln k_0$.  This approximation is well justified (see Sec.  
\ref{IV}) in view of the smallness of      $\partial^2_{\cal E}  \ln k_0$, compared to other ingredients of Eqs. 
(\ref{R1})-(\ref{R3}).  The   value of the $\alpha^3$  QED recoil correction $\alpha_{\rm d}^{(3,1)}$ obtained by us 
equals to  0.01112(1)$\cdot 10^{-6}a_0^3$ and is only about nine times smaller than the $\alpha^2$ relativistic recoil 
correction $\alpha_{\rm d}^{(2,1)}$.  It may be of interest to note that the relative magnitude of the finite mass 
contributions to the nonrelativistic,  
$\alpha_{\rm d}^{(0)}$, relativistic, $\alpha_{\rm d}^{(2)}$,  and QED, $\alpha_{\rm d}^{(3)}$,  components of the  
static polarizability of helium-4 are quite different. Specifically, we found that
\  $\alpha_{\rm d}^{(0,1)}/\alpha_{\rm d}^{(0)} \approx 3.2/M$,
\  $\alpha_{\rm d}^{(2,1)}/\alpha_{\rm d}^{(2)} \approx 8.5/M$, and  
 $\alpha_{\rm d}^{(3,1)}/\alpha_{\rm d}^{(3,0)} \approx 2.7/M$. 

In Table \ref{BD2}, the results of our calculations are added to the data obtained in earlier work \cite{Piszczatowski:15,Puchalski:16}  and  compared 
with the most recent experimental determination \cite{Gaiser:18} of $\alpha_{\rm d}$, given in terms  of the molar 
polarizability $A_{\epsilon}$=$ {4\pi}{ }\alpha_{\rm d}N_A/3$.  The agreement between theory and experiment is very 
good, although the uncertainty of the experimental value is an order of magnitude larger than that of the theoretical 
determination. This high theoretical accuracy appears to be presently sufficient for metrological purposes  
\cite{Hendricks:18,Gavioso:16,Jousten:17}.  As shown in Table  \ref{BD2},  this accuracy is currently limited by the 
incomplete calculation of the $ \alpha^4$  QED correction. Complete calculations of this correction for the energy 
levels of helium have  been  very challenging \cite{Pachucki:06a,Pachucki:06b,Yerokhin:10} and  have  not been attempted 
when the effect of the interaction with external electric field is included in the Hamiltonian. The recent successful 
calculation  of the $\alpha^4$   QED correction for the hydrogen molecule \cite{Puchalski:16a} shows that a similar 
calculation for the helium  atom  in the uniform electric field, a system of the same symmetry as H$_2$, may  be 
possible if   accuracy higher than achieved in the present paper is required for metrological or other applications.

 \section*{Appendix  }

In this Appendix we present derivation of the second electric-field  derivative of  the  $C_3$ coefficient  that determines the $\omega^{-3}$  term (equal to $4{D}C_3\,\omega^{-3}$)  in the large-$\k$ asymptotic expansion of $J(\omega)$.  To obtain this expansion, we consider an  auxiliary function  $\bphi$ defined by 
\begin{align}\label{defphi}
   \left(H-E + \omega \right) \bphi = \bps\psi. 
\end{align}
where   $\psi$ is the real ground-state eigenfunction of the  Hamiltonian 
 $H$=$H_0+{\cal E} z$.  For the sake of brevity,  in Eq. (\ref{defphi}) and below we suppress the 
dependence of $\bphi$ on ${\cal E}$ and on $\omega$. 
Obviously $J(\omega)=\langle \bps\psi  | \bphi\rangle$,  but  it is advantageous to compute  $J(\omega)$ from the expression
\begin{align}\label{J1}
 J (\omega) =  \langle\bphi|\bps\psi \rangle  +  \langle\psi|\bps\bphi \rangle  -  \langle \bphi | 
H-E + \omega  |\bphi \rangle,
\end{align}
which for an approximate $\bphi$ gives $J (\omega) $ with an error quadratic in the error of $\bphi$ [and provides a lower bound 
to $J(\k)$].  
%
Following Schwartz  \cite{Schwartz:61},  we  write  $\bphi$ in the form  
\begin{align}
\label{psi1as}
 \bphi= \frac{1}{\omega }\,\bps\psi  +i\, \bU,
\end{align}
where the  real function $\bU$ collects  terms that vanish  faster than $\omega^{-1}$.   Inserting Eq. (\ref{psi1as}) 
into  Eqs (\ref{defphi}) 
we find  that $\bU$  
  obeys the relation
\begin{align}
\label{eqU}
 \left(H-E + \omega \right)\bU = -  \frac{2}{\omega }  \ba\, \psi -\frac{2}{\omega} \,{\cal E} \bk\, \psi.
\end{align}
and that $J(\omega)$ can be represented in the form 
\begin{align}
\label{jj}
 J(\omega)  = \frac{1}{\k} \langle\psi\vert\bp^2\psi\rangle - \frac{D}{\omega^2 } -
 \frac{4}{\k}\,\langle \, \ba\, \psi|\bU\rangle - \frac{4}{\k} \,{\cal E}\,
 \langle\psi|\bU\rangle \,\bk 
 &- \,\langle \bU|H-E_0+ \omega |\bU\rangle,\end{align}
 where $\bk$ is the unit vector on the $z$ axis and 
 $ \ba =    {\br}_1r_1^{-3} +    {\br}_2r_2^{-3}$, 
so that $[H,\bp]=2i(\ba-  {\cal E}\bk)$.  

It is obvious that the solution of Eq. (\ref{eqU}) can be written as $\bU =\bU_1 +\bU_2$, where 
\begin{equation} \label{U2}
\bU_2 =- \frac{2}{\omega^2} \,{\cal E} \bk\, \psi
\end{equation} 
and $\bU_1$ is the solution of Eq. (\ref{eqU}) with the last term neglected.   Schwartz \cite{Schwartz:61} has found an approximate solution for 
$\bU_1$ which, when inserted in Eq. (\ref{jj}),  correctly recovers the $\k^{-5/2}$, $\k^{-3}\ln\k$, and $\k^{-3}$ terms in the large-$\k$ asymptotic expansion of  $J(\k)$. His result is \cite{Schwartz:61}
\begin{align}\label{U1}
   \bU_1 =- \frac{2}{\omega ^2}\sum_i\frac{\br_i}{r_i^3}\left[1-e^{-\mu r_i}(1+\mu 
r_i)\right]\psi,
 \end{align}
where  $\mu=(2\k)^{1/2}$. In deriving Eq. (\ref{U1}),  Schwartz neglected the potential energy  terms in the Hamiltonian 
on the left-hand side
 of Eq. (\ref{eqU}) (see Ref. \onlinecite{Bukowski:92} for an alternative derivation   based on this assumption).  
Thus, Eq. (\ref{U1}) is  valid
also  for an atom in the electric field  that  enters $\bU_1$ only through the field dependence of  $\psi$. 

Combining  Eqs.~(\ref{jj})-(\ref{U1}), we find after some   cancellations that 
\begin{align}
\label{jjj}
 J(\omega)  = \frac{1}{\k}\langle\psi\vert\bp^2\psi\rangle - \frac{D}{\omega^2 }  +  J_1(\omega) + J_2(\omega)  +{\cal O}(\k^{-7/2}),
\end{align}
where
\begin{equation} \label{J1b}
J_1(\k) = -  \frac{4}{\k}\,\langle \, \ba\, \psi|\bU_1\rangle  
- \,\langle \bU_1|H-E_0+ \omega |\bU_1\rangle
\end{equation} 
and 
\begin{equation} \label{J2}
J_2(\k) = - \frac{4}{\k} \langle  \ba  \psi|\bU_2\rangle - \frac{4}{\k}  {\cal E} 
 \langle\psi|\bU_2\rangle \,\bk 
 -  \langle \bU_2|H-E_0+ \omega |\bU_2\rangle = \frac{4}{\k^3}\, {\cal E}^2 +\frac{8}{\k^3} \, \langle\psi\vert a_z \psi\rangle\, {\cal E}
 ,
\end{equation} 
with $a_z$=$\ba\bk$. 
Derivation of the large-$\k$ expansion of  $J_1(\k)$ is complicated. It has been performed through the $\k^{-3}$ term by Schwartz \cite{Schwartz:61}. His result, confirmed by Forrey and Hill~\cite{Forrey:93},  is
\begin{align}
\label{J1a}
 J_1(\k)  = \frac{4\sqrt{2}{D}}{\omega^{5/2}} -    \frac{8{D}}{\k^3} \ln\omega + \frac{4{D}}{\k^3}C^{(1)}_3
+{\cal O}(\k^{-7/2}).
\end{align}
where the   coefficient $C^{(1)}_3$, depending on ${\cal E}$ via $\psi$, is given by the expression 
\cite{Schwartz:61,Forrey:93} 
 \begin{equation}\label{C31}
C^{(1)}_3  =   
      4\left(\frac{1}{2}\ln{2}-\frac{1}{2}-\gamma\right)
       -\frac{1}{ D} \int_0^\infty\ln{r}\frac{d^2 \bar{\rho}(r)}{d r^2}dr
     +\frac{2}{D}\, \left<\psi  | \br_1\br_2 r_1^{-3}r_2^{-3} \psi  \right>     ,
\end{equation}
with  $\bar{\rho}(r)$ denoting   the angular average of the electron density
 $\rho(\br)$$=$$\langle \psi  |\delta(\br$$-$$\br_1)\! +\!\delta(\br$$-$$\br_2) | \psi  \rangle$.
 
From Eqs. (\ref{jjj}), (\ref{J2}), and (\ref{J1a}), it is clear that $C_3 =C^{(1)}_3+C^{(2)}_3 $, where $C^{(2)}_3$  
is the contributions from $J_2(\k)$ given by 
 \begin{equation}\label{C32}
C^{(2)}_3  = \frac{1}{D} {\cal E}^2  +\frac{2}{D} \langle\psi\vert a_z \psi\rangle  {\cal E}  . 
\end{equation}

  Calculating the second electric-field derivative  at ${\cal E} =0$  we arrive at
 \begin{equation}\label{dC32}
\partial^2_{\cal E}C^{(2)}_3  = \frac{2}{D}       -\frac{8}{D} \langle\psi_0\vert z R_0 a_z\psi_0\rangle. 
\end{equation}
Since $\langle\psi_0\vert z R_0 a_z\psi_0\rangle=1/2$,  we finally obtain (cf. Table \ref{I})
\begin{equation}\label{dC32fin}
\partial^2_{\cal E}C^{(2)}_3  = -\frac{2}{D} = -      0.04395503(1).
\end{equation}

The numerical evaluation of $\partial_{\cal E}C^{(1)}_3$ is much more difficult.  Performing electric-field differentiation of 
Eq. (\ref{C31}), one finds
 \begin{equation}\label{d2C3}
  \partial_{\cal E}C^{(1)}_3 = -
\frac{1}{   D}\left(   \partial^2_{\cal E} I_1   -  \frac{ I_1 }{  D}  \,\partial^2_{\cal E} {  D}\right)   
+
\frac{2}{  D}\left( \partial^2_{\cal E} I_2-  \frac{ I_2 }{  D}  \,\partial^2_{\cal E} {  D}\right),   
 \end{equation}
where  $I_1$ and $I_2$ are the integrals 
\begin{equation}\label{Int1}
I_1 = \int_0^\infty\ln{r}\frac{d^2 \bar{\rho}(r)}{d r^2}dr \ \ \end{equation}
and
\begin{equation}\label{Int2}
I_2 = \left<\psi\vert \br_1\br_2 r_1^{-3}r_2^{-3} \vert\psi \right>.
\end{equation}
%
%
 The electric-field derivatives 
  $\partial^2_{\cal E}I_2$ and   $\partial^2_{\cal E} {\rho}(\br)$,  needed for the evaluation 
of $\partial^2_{\cal E}C_3^{(1)}$   via Eqs.~(\ref{d2C3}) and (\ref{Int1}),  were computed using Eq. (\ref{E21}) and basis sets with 
$K_0$ = 128, 256, and 512 optimized as described in Sec. \ref{II}.  The convergence of calculations was rather slow and  we found that   $\partial_{\cal E}^2 C_3^{(1)}=-0.00834(2)$.  The same value was obtained using an alternative formula  
for $C_3^{(1)}$ in which the last two terms in Eq. (\ref{C31}) are replaced by the finite part  of 
$\left(\psi\vert\ba^2\psi\right)/  {D}$ [cf. Eq. (19) in Ref. \onlinecite{Schwartz:61}].   One may note that  the 
contribution $\partial_{\cal E}^2 C_3^{(2)}$,  
derived in the present  paper,  is about five times larger in absolute value than  the second 
electric-field derivative of the formula for  $C_3$ given in Ref. \onlinecite{Schwartz:61}.

Adding up $\partial_{\cal E}^2 C_3^{(1)}$ and  $\partial_{\cal E}^2 C_3^{(2)}$,  we finally find that  $\partial_{\cal E}^2 C_3  =  -0.052\,30(2)$.   This value compares  reasonably  well with the   value $-0.053(1)$
obtained from fitting  the derivative of $t^{-3} \dk f(t)$ at $t=0$ [cf. Eqs. (12) and (15)].

\begin{acknowledgments}
We thank Grzegorz \L ach and Krzysztof Pachucki for useful discussions.   
This project (QuantumPascal project 18SIB04) has received funding from
the EMPIR programme co-financed by the Participating States and from the
European Union's Horizon 2020 research and innovation program.
B.J. and M.L., and M.P. acknowledge support by the National 
Science Center, Poland within the Project No. 2017/27/B/ST4/02739.    
The support of the NSF Grant No. CHE-1900551 is also acknowledged.  
\end{acknowledgments}

   \bibliography{BL_REFs}

  \end{document}